\documentclass[structabstract]{aa}
\usepackage{graphicx}
\usepackage{txfonts}
\usepackage{rotating}
\usepackage{lscape}
\usepackage{longtable}
\usepackage{natbib}
\bibpunct{(}{)}{;}{a}{}{,}

\begin{document}

\title{Star formation in two irradiated globules around Cygnus~OB2
\thanks{Based on observations collected at the Centro Astron\'omico Hispano en Andaluc\'\i a (CAHA) at Calar Alto, operated jointly by the Junta de Andaluc\'\i a and the Instituto de Astrof\'\i sica de Andaluc\'\i a (CSIC); with the IAC80 optical telescope operated on the island of Tenerife by the Instituto de Astrof\'\i sica de Canarias in the Spanish Observatorio del Teide; and with the Italian Telescopio Nazionale Galileo (TNG) operated on the island of La Palma by the Fundaci\'on Galileo Galilei of the INAF (Istituto Nazionale di Astrofisica) at the Spanish Observatorio del Roque de los Muchachos of the Instituto de Astrof\'\i sica de Canarias.}
}
\author{F. Comer\'on\inst{1}
\and N. Schneider\inst{2}
\and A.A. Djupvik\inst{3,4}
}
 \institute{
  European Southern Observatory, Karl-Schwarzschild-Str. 2, D-85748 Garching bei M\"unchen, Germany\\
  \email{fcomeron@eso.org}
  \and
   I. Physik. Institut, University of Cologne, Z\"ulpicher Str. 77, D-50937 Cologne, Germany
  \and
  Nordic Optical Telescope, Rambla Jos\'e Ana Fern\'andez P\'erez, 7, E-38711 Breña Baja, Spain
  \and
  Department of Physics and Astronomy, Aarhus University, Munkegade 120, DK-8000 Aarhus C, Denmark
  }
%
%
\date{Received; accepted}
\abstract
{The ultraviolet irradiation and the action of stellar winds of newly formed massive stars on their parental molecular environment often produces isolated or small clouds, some of which become sites of star formation themselves.}
{We investigate the young stellar populations associated with DR~18 and ECX~6-21, which are two isolated globules irradiated by the O-type stars of the Cygnus~OB2 association. Both are HII regions containing obvious tracers of recent and ongoing star formation. We also study smaller isolated molecular structures in their surroundings.}
{We combined near-infrared broad- and narrow-band imaging with broad-band imaging in the visible and with archive images obtained with the Spitzer Space Telescope. We used the joint photometry to select young stellar objects (YSOs), simultaneously estimating their intrinsic properties and classifying them according to the characteristics of their infrared excess. We also present low-resolution visible spectroscopy of selected sources.}
{We reproduce previous findings of an extended population of YSOs around both globules, dominated by the more evolved classes, associated with the general Cygnus~OB2 population. Both globules contain their own embedded populations, with a higher fraction of the less-evolved classes. Masses and temperatures are estimated under the assumption of a common age of 1~Myr, which has been found to appropriately represent the general Cygnus~OB2 YSO population but is most probably an overestimate for both globules, especially ECX~6-21. The early-B star responsible for the erosion of DR~18 is found to be part of a small aggregate of intermediate-mass stars still embedded in the cloud, which probably contains a second site of recent star formation, also with intermediate-mass stars. We confirm the two main star forming sites embedded in ECX~6-21 described in previous works, with the southern site being more evolved than the northern site. We also discuss the small globule ECX~6-21-W ($= G79.8+1.2$), and propose that its non thermal radio spectrum is due to synchrotron emission from an embedded jet, whose existence is suggested by our observations.} 
{The extreme youth of some of the YSOs suggests that star formation in both globules started after they became externally irradiated. The populations of both globules are not found to be particularly rich, but they contain stars with estimated masses similar or above that of the Sun in numbers that hint at some differences with respect to the star formation process taking place in more quiescent regions where low-mass stars dominate, which deeper observations may confirm.}



\maketitle

\section{Introduction \label{intro}}

Globules, pillars, and proplyd-like clumps are a natural result of the erosion of molecular clouds being exposed to the destructive action of the ultraviolet light and winds from nearby hot, massive stars \citep[][and references therein]{Schneps80,Schneider16, Goicoechea20}. Clumps in the denser parts of those clouds take a longer time to be evaporated, and in the process they usually evolve into isolated or nearly isolated structures whose morphology and physical conditions are partly determined by the harsh environment in which they find themselves.

These structures, which we generically refer to as globules, are sometimes found to harbor star forming sites \citep{Hawarden76,Reipurth83,Rebull11,Wright12,Kun17,Schneider12,Schneider21,Djupvik17,Reiter20}. This naturally raises questions on the star formation process in such conditions. Examples of matters to investigate concern whether star formation was taking place in them before their exposure to external irradiation, or was instead triggered by the exposure; whether the collective properties of the stars that formed in irradiated globules, such as their mass function, are the same as those of aggregates that formed in more quiescent conditions; whether stars tend to form near the edges of globules, where the effects of radiation, ionization fronts, and stellar winds are most immediate, or rather in their dense interiors less affected by the external conditions. Since the ultimate destruction of their parental molecular clouds is a natural outcome of the formation of massive stars, and as the formation of transient globules is a byproduct of such destruction, irradiated globules may be a relatively common environment of star formation.

The Cygnus~X molecular complex \citep{Schneider06}, physically linked to the very massive association Cygnus~OB2, offers a fertile ground for the study of star formation in the presence of massive stars at the relatively nearby distance of 1.5~kpc \citep{Rygl12, Comeron20}. Compact thermal radio continuum sources were first identified in  Cygnus X by \citet{Downes66}, and they have been revealed in increasing detail at radio, submillimeter, and infrared wavelengths by subsequent surveys \citep[e.g.,][]{Landecker84,Wendker84,Wendker91,Schneider07,Beerer10}. Sensitive mid-infrared images such as those obtained with the Spitzer Space Telescope show that many of those regions can be associated with globules whose shapes and brightness distributions indicate that they are illuminated and eroded by the combined action of the O stars of Cygnus~OB2 \citep{Wright12}. The estimated radiation intensity at their position, together with their derived masses and densities, may be used to estimate the time remaining until their complete evaporation \citep{Schneider16}, which can be accelerated by the formation of massive stars in their interiors.

In this paper we present new observations and discuss the properties of the stellar content of two of these globules, DR~18 and ECX~6-21, and of some objects in their proximity probably related to them. Previous works have considered the overall distribution and collective properties of young stellar objects in the Cygnus~OB2 / Cygnus~X region \citep{Guarcello13,Kryukova14}, the far-infrared and radio properties of globules and related structures \citep{Schneider16,Isequilla19} and the properties of selected proplyd-like structures with embedded stars \citep{Wright12,Guarcello14}. Complementary to those studies, here we focus on the star formation sites of both globules,  carefully discarding spurious detections resulting from small bright spots in the structured nebulosity. The properties of the young stellar objects most likely associated with them, and the possible impact of their most massive stars are discussed. This provides a necessary basis for further investigations of these globules leading to the modelling of the role of their young stellar population on the properties of their interstellar gas and the relative contributions of external and internal radiative and mechanical input.

\section{DR18 and ECX6-21 \label{DR18ECX621}}

\begin{figure*}[ht]
\begin{center}
\hspace{-0.5cm}
\includegraphics [width=14cm, angle={0}]{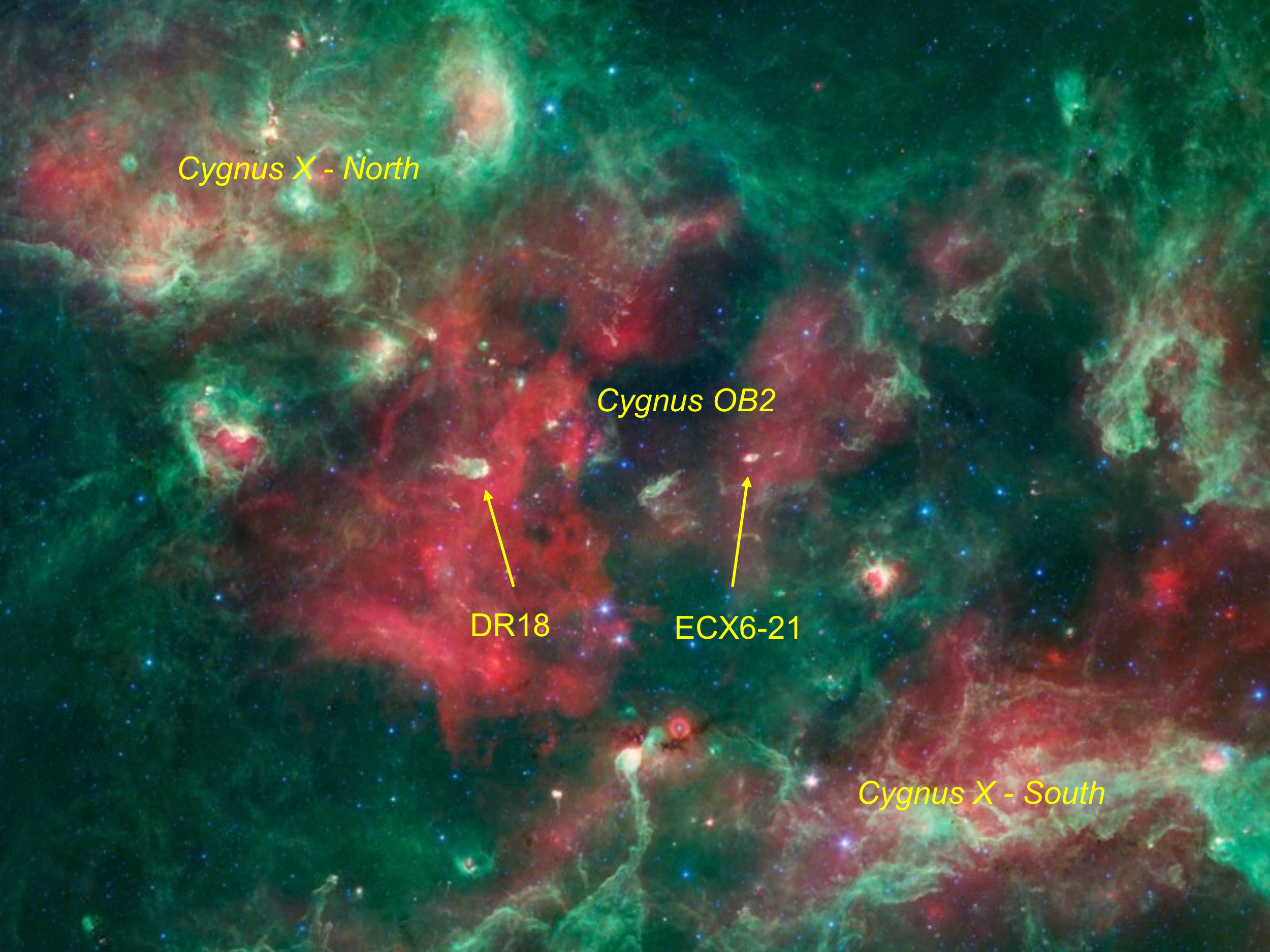}
\caption []{Panoramic view of the Cygnus~OB2 / Cygnus~X complex obtained by the Spitzer Space Telescope, with our two targets indicated. The blue, green and red channels correspond to emission in the 3.6, 5.8, and 24~$\mu$m bands. The area covered is $4^\circ 6 \times 3^\circ 2$. The image highlights the distribution of warm dust in the region around the stars of Cygnus~OB2, which is located near the center of the image and has a radius $\sim 1^\circ$. The image clearly shows the effect of the erosion produced by the O stars in Cygnus~OB2 on the surrounding interstear medium, producing a cavity where some isolated globules,among them DR~18 and ECX~6-21, persist. The images were obtained for the Cygnus-X Spitzer Legacy Survey \citep{Beerer10}.}
\label{Cyg_Spitzer}
\end{center}
\end{figure*}

DR~18 and ECX~6-21 are two isolated globules projected on the central region of Cygnus~OB2 in the cavity that surrounds it and that is roughly delimited by the molecular complexes of Cygnus~X-North and South \citep{Schneider06}; see Figure~\ref{Cyg_Spitzer}. Their approximate angular sizes are respectively $10'0 \times 4'5$ and $3'0 \times 1'5$, which correspond to $4.4 \times 2.0$~pc$^2$ and $1.3 \times 0.7$~pc$^2$ at the adopted distance of 1.5~kpc.

 Globules and related structures are more abundant near the edges of the cavity where they are in the process of becoming detached from the main molecular complex \citep{Beerer10}. This also makes the study of their embedded populations more difficult due to confusion, which is mitigated thanks to the relative isolation in the case of DR~18 and ECX~6-21, although YSOs belonging to the general population of Cygnus~OB2 pervade the whole area \citep{Guarcello13}. The velocity of the molecular gas associated with the globules clearly associates it with the Cygnus~X molecular complex \citep{Schneider06}, showing that they are not line-of-sight features. 

\begin{figure*}[ht]
\begin{center}
\hspace{-0.5cm}
\includegraphics [width=12cm, angle={0}]{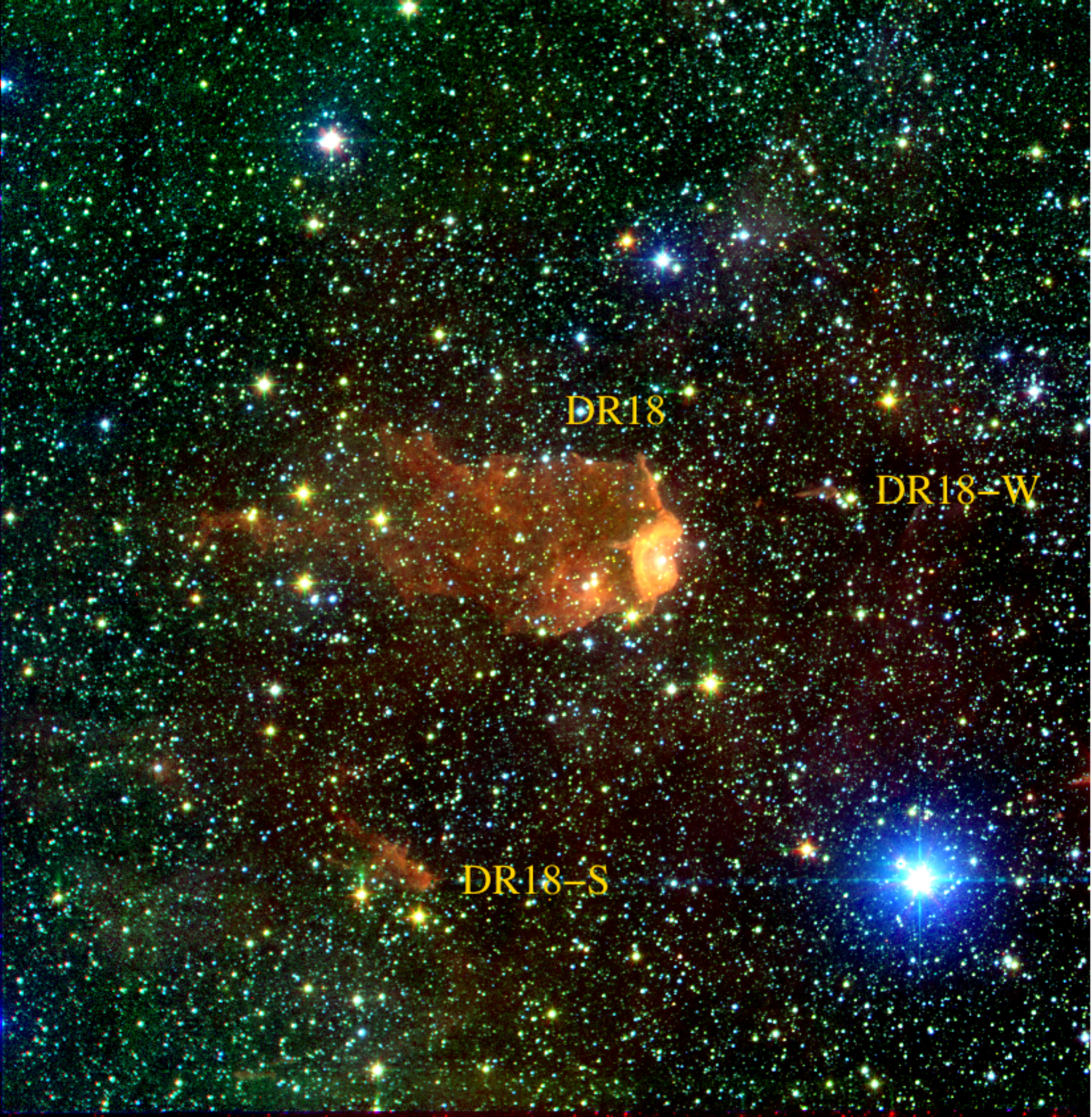}
\caption []{Red-Green-Blue color composite of the field around DR~18 covered by our observations. The blue and green channels correspond respectively to the $J$ (1.25~$\mu$m) and $K_S$ (2.16$\mu$m) bands, generated from the new observations presented in this work. The red channel is the Spitzer image in the $I2$ (4.5~$\mu$m) IRAC band. The field covered is $24'7 \times 25'3$.}
\label{DR18_JKsI2}
\end{center}
\end{figure*}

DR~18 has been included in many surveys, but to our knowledge the only dedicated study is still the one by \citet{Comeron99} at visible and near-infrared wavelengths. The main feature of DR~18 is an arc-shaped nebula, bright in the infrared, on its western side whose shape is clearly due to the local erosion by an early B-type star. The erosion of the globule generates a stream of ionized gas in the direction of the star, faintly visible in H$\alpha$ images, producing what appears to be a small bow shock ahead of the star when the stream of gas encounters the stellar wind. The bright arc-shaped nebula marks the photodissociation region (PDR) at the interface between the cloud and the ionization front. Near-infrared images of DR~18 show faint emission extending to the east of the arc-shaped nebula, but the actual size and structure of the globule are revealed by the Spitzer images longward of $\sim 3$~$\mu$m (Figure~\ref{DR18_JKsI2}). The western half of DR~18 is at the edge of the region mapped with the Giant Metrewave Radio Telescope by \cite{Isequilla19}, showing the contour of the cloud and a strong brightening around the position of the B star and the PDR. A nearby smaller globule located about $4'$ west of the head is discussed by \citet{Wright12}, \citet{Schneider16} and \citet{Isequilla19} as one of their evaporating proplyd-like features in the area. We refer to it here as DR~18-W. Another smaller nebula, hereafter DR~18-S, appears $8.5'$ south of the main body of DR~18. All three nebulae are elongated in the east-west direction and display the head-tail structure typical of many globules, with the head on the west facing the center of Cygnus~OB2 and the more diffuse tail extending eastward.

\begin{figure*}[ht]
\begin{center}
\hspace{-0.5cm}
\includegraphics [width=12cm, angle={0}]{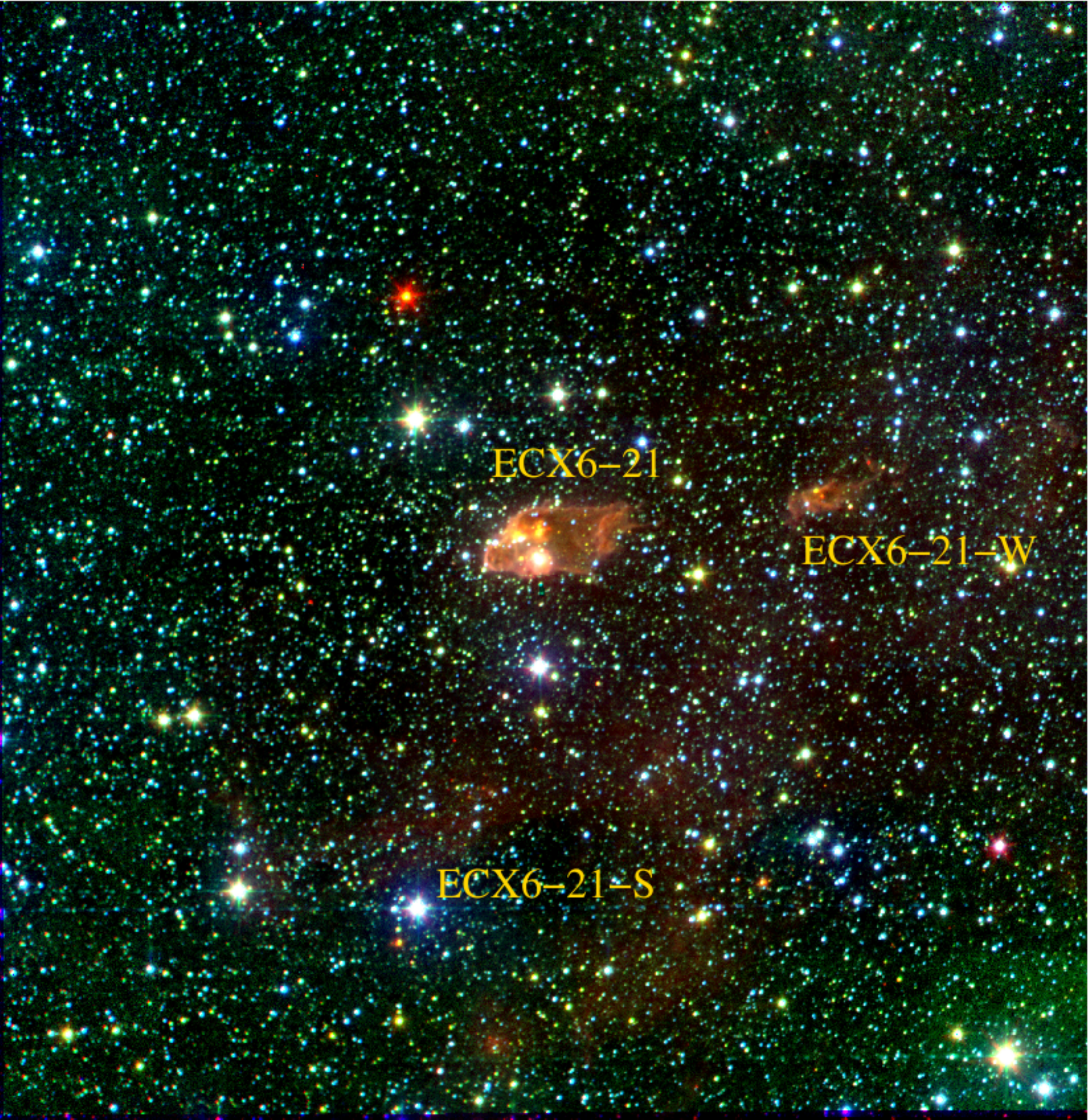}
\caption []{Same as Fig.~\ref{DR18_JKsI2}, but centered around ECX~6-21. The field is now $20.7 \times 21'3$.}
\label{ECX6-21_JKsI2}
\end{center}
\end{figure*}

Similar to DR~18, ECX~6-21 (Fig.~\ref{ECX6-21_JKsI2}) was first identified as a compact thermal source \citep{Wendker84,Wendker91}. Although it is within the boundaries of the DR~11 HII region as defined in the low angular resolution survey of \citet{Downes66}, carried out with a beam size of $10'$, the central position given by those authors for DR~11 is roughly centered on the position of an extended nebulosity $7'$ south of ECX~6-21, which we consider as a distinct structure referred to as ECX~6-21-S in this work. The eastern side of ECX~6-21 appears in visible images as a sharply delineated thin crescent of H$\alpha$ emission, and as in the case of DR~18 the true extent of the globule is outlined by infrared observations. The VLA high-resolution 6~cm continuum map presented by \citet{McCutcheon91} shows that the ionized gas emission traces well the H$\alpha$ emission and it extends outward of the globule.

In contrast with DR~18, the smaller globule ECX~6-21 has been the subject of several dedicated studies, mostly at far-infrared and radio wavelengths. The existence of active star formation in it is revealed by water masers and NH$_3$ emission \citep{Palla91,Brand94,Molinari96,Urquhart11}. \citet{Zhang05} reported the existence of a CO outflow, and \citet{Varricatt10} noted the existence of jet-like features in narrow-band H$_2$ images associated with bright near-infrared sources. There is only one compact radio continuum source embedded in the cloud, whose position is coincident with the origin of the H$_2$ outflows detected by \citet{Varricatt10}. This seems to be the only embedded object in the globule able to significantly ionize its surroundings. Near-infrared images show a loose cluster of YSOs, which was discussed by \citet{Kumar06}, in which \citet{Guarcello13} note a very high fraction of Class I over Class II sources. Three concentrations of YSOs stand out  coincident with dust temperature peaks \citep{Verma03,Ramachandran17}. Masses, ages and accretion rates of most YSOs have been estimated by \citet{Ramachandran17}, who find them to be very young, with ages below 1~Myr and with temperatures below 10,000~K, including the YSO located at the position of the radio continuum point source.

Some $6'$ west of ECX~6-21 there is a smaller globule with a head-tail structure, which we call hereafter ECX~6-21-W. There appears to be some confusion in the literature as to the nature of this object. Its position is coincident with that of the source G79.8+1.2, which has been classified as a supernova remnant. This object was included the catalog of non-thermal radiosources of \citet{Milne70}, where it is nevertheless listed as an extended source with a size of $13' \times 20'$, far larger than its dimensions ($0'5 \times 1'0$) in infrared images. In fact, \citet{Milne70} identifies G79.8+1.2 with DR~11, which as noted above should probably be regarded as a conglomerate of distinct sources. Both hydrogen recombination lines \citep{Dickel72} and optical line ratios \citep{Sabbadin76} at the nominal position of ECX~6-21-W clearly classify it as a HII region. However, newer radio continuum observations at 1400 MHz and 350 MHz, now with a synthesized beam size comparable to the extent of ECX~6-21-W, find a non-thermal spectrum at this position \citep{Setia03}. Sources with non-thermal spectral indices are not extremely rare in star forming regions, and \citet{Rodriguez93} have considered the conditions under which inhomogeneous thermal sources may disguise as non-thermal. They conclude that only extreme density conditions, unlikely to remain stable against gravitational collapse, may produce such effect, favoring instead genuine synchrotron emission as the cause of such indices. Our observations yield some insight and a possible solution to this apparent conflict, which we discuss in Section~\ref{ECX6-21_around}.

\section{Observations\label{observations}}

\subsection{Near-infrared imaging \label{ir_imaging}}

We obtained near-infrared images of DR~18 and ECX~6-21 and their respective environments using the PANIC wide field near infrared camera at the Calar Alto 2.2m telescope \citep{Baumeister08} on 4 nights between 4-5 and 7-8 August 2016. At the time of the observations the camera had a mosaic of $2 \times 2$ detectors of $2048 \times 2048$~pixels$^2$ with a pixel scale of $0''450$ per pixel, which has since then been replaced by a single $4096 \times 4096$ detector due to the very large number of bad pixels of three of the former detectors. Our results are based on the field imaged by the least affected detector, which still yielded a field of view of $15' \times 15'$. We used the $J$, $H$, and $K_S$ broad-band filters, as well as narrow-band filters centered on the wavelengths of 2.122~$\mu$m and 2.166~$\mu$m isolating emission in the H$_2 \ S(1 \rightarrow 0)$ and Br$\gamma$ lines respectively. Each image was built from the combination of frames on a dithered pattern of 9 telescope pointings on a $3 \times 3$ square grid. The grid size was chosen to be about twice the angular size of each globule, being $5' \times 5'$ for ECX~6-21 and $9' \times 9'$ for DR~18. The observations of DR~18 and ECX~6-21 were interleaved with those for another project that used the same setup and data reduction procedure and extraction of point-spread function (PSF) photometry, which are described in \citet{Comeron19}. The reader is referred to the details provided on the data reduction in Section 2 of that work, as they are entirely applicable to the infrared imaging observations presented here. The $5\sigma$ detection limits are essentially identical to those estimated in that work, $J \simeq 19.5$, $H \simeq 18.0$, $K_S \simeq 16.5$.

We carried out an approximate flux calibration for the Br$\gamma$ and H$_2$ narrow-band filters by assuming that on the average $[{\rm H}_2] - K_s = 0$ and $[{\rm Br}\gamma] - K_S = 0$ for the vast majority of stars in the field, where $[{\rm H}_2]$ and $[{\rm Br}\gamma]$ is the magnitude in each of those bands, and derived in this way the corresponding approximate zeropoints. Deviations from the average are common for individual stars, and depend on the spectral energy distribution in the $K-$band, the reddening, and the presence and strength of emission and absorption lines. Assuming that only a strong emission or absorption feature causes a non zero difference between the band $f$ containing the feature and the continuum sampled by the band $c$, the equivalent width of the feature, $EW(f)$, can be estimated as

\begin{equation}
EW(f) = W(f) [10^{-0.4 ([f]-[c])}-1]  \label{ew}
\end{equation}

\noindent where $W(f)$ is the passband of the filter containing the feature, and $[f]$ and $[c]$ are the magnitudes in the narrow-band filter containing the feature and the filter sampling the continuum. According to the filter specifications of the PANIC camera, $W({\rm Br}\gamma) = 550$\AA \ and $W({\rm H}_2) = 340$\AA .

\subsection{Visible imaging \label{vis_imaging}}

To extend the determination of the spectral energy distributions of stars with low-to-moderate extinction, complementary observations in the SDSS $g'r'i'$ bands centered on DR~18, ECX~6-21 and their respective immediate fields, covering an area of $10' \times 10'$ around each source, were obtained using the CAMELOT imager at the 0.8m IAC80 telescope of the Teide Observatory (Canary Islands, Spain). The observations were spread over five nights, from 15 to 19 September 2016. Integration times amounted to 6000, 300, and 100 seconds in the $g'$, $r'$, and $i'$ bands respectively, reaching $5\sigma$ detection limits $g' \sim 21.5$, $r' \sim 21.0$, $i' \sim 20.5$. As in the case of the near-infrared images, PSF photometry was performed on the images. Color coefficients and photometric zeropoints were calculated taking as reference the magnitudes of stars in common listed in the UCAC4 catalog \citep{Zacharias13}.

\subsection{Archival Spitzer imaging \label{spitzer}}

Calibrated Post-BCD images covering the area imaged with our PANIC observations obtained with the IRAC instrument onboard the Spitzer Space Telescope in the I1 (3.6~$\mu$m), I2 (4.5~$\mu$m), I3 (5.8~$\mu$m), and I4 (8.0~$\mu$m) bands were identified in the Spitzer archive, downloaded, and combined into a single image for each band. All the images used here were obtained for the Cygnus-X Spitzer Legacy Survey \citep{Beerer10}, whose Principal Investigator was J.~Hora, and they have been used in many other studies of the region. The DR~18 area was covered by the Spitzer Astronomical Observation Requests (AOR) numbers 22499584, 22499840, 27106048, 27107584, and 27112704, and the ECX~6-21 area by AORs numbered 22500096, 27111936, and 27112704.

PSF instrumental photometry was performed in the same way as for the near-infrared and visible images. Since Spitzer Post-BCD images are already flux-calibrated, magnitudes could be directly obtained from the combined images using the corresponding flux zeropoints for the IRAC bands. We did not use the images obtained at longer wavelengths with the MIPS instrument, due to confusion with all-pervading bright nebulosity in both globules and saturation in some areas.

\subsection{Visible spectroscopy \label{spectroscopy}}

We selected some sources of interest, which we discuss below, for spectroscopic follow-up in the visible. The criteria for selection of the sources to be spectroscopically observed were their detectability in the $r'$ band images, the evidence for infrared excess, and their location in or near the globules. Four sources in the DR~18 main cloud were observed using the CAFOS instrument at the Calar Alto 2.2m telescope (Andalusia, Spain) on 7 and 8 August 2017, and on 27 and 28 July 2021. The grating used produced spectra spanning the wavelength range from 6300 to 8800~\AA , although the very red colors of all the stars observed made the blue end of the spectra unusable for spectral classification purposes. The resolution was $\lambda / \Delta \lambda \simeq 700$ with the $1''2$-wide slit used, and the total exposure time per target was 60 minutes. Additionally, a spectrum of the only YSO identified in DR~18-W was obtained with the DOLORES spectrograph at the Telescopio Nazionale Galileo (TNG) at the Observatory of Roque de los Muchachos (La Palma, Canary Islands, Spain). The Low-Resolution Red grism was used, with a wavelength coverage overlapping that of CAFOS and at a very similar spectral resolution. The exposure time of this latter spectrum was also 60 minutes. 

The spectra were normalized to the continuum, and classified based on the atlas of low-resolution stellar spectra in the near-infrared of \citet{Torres93}. The instrument used, observation date, and our spectral classification of the observed sources are summarized in Table~\ref{spectra_list}.

\begin{table}
\caption{Objects for which spectral classifications were obtained\label{spectra_list}}
\begin{tabular}{ccccccccc}
\hline\hline
\noalign{\smallskip}
Star & Instrument & Date & Sp. type \\
\noalign{\smallskip}\hline\noalign{\smallskip}
DR 18-05 & CAFOS & 7 Aug 2017 & B2 \\
DR 18-14 & CAFOS & 28 Jul 2021 & $>$F5:  \\
DR 18-15 & CAFOS      & 28 Jul 2021 & early K   \\
DR 18-16 & CAFOS      & 28 Jul 2021 & $>$F5e   \\
DR 18-W-01 & DOLORES & 11 Oct 2021 & ?:e \\
ECX 6-21-03 & CAFOS & 29 Jul 2021 & mid-A \\
ECX 6-21-14 & CAFOS & 7 Aug 2017  & B2   \\
\hline                            
\end{tabular}        
\smallskip\\
\end{table}                      

\section{Selection of young stellar objects \label{ysosel}}

A first list of sources detected in each band was obtained by applying the DAOFIND task \citep{Stetson87} layered on IRAF on the reduced images in each band. Source detection in some regions is complicated by the presence of bright and highly structured nebulosity, leading to spurious detections in each band. Many extended or dubious sources were automatically excluded by requiring both the full-width at half-maximum and the ellipticity of their PSF to differ less than 30\% from those of relatively isolated and bright point sources. Spurious detections were largely removed from our source catalog by requiring detection in at least three of the $g', r', i', J, H, K_S, I1, I2, I3, I4$ bands used in our observations. In this way, we obtained a catalog of 25,339 sources in the field of DR~18, and 15,061 sources in the field of ECX~6-21. The vast majority of those sources are unrelated foreground and background stars, obscured by widely varying degrees of interstellar extinction.

A number of criteria based on the combination of near- and mid-infrared colors have been proposed for the selection of the various classes of infrared excess sources \cite[e.g.,][]{Megeath04,Robitaille06,Evans09}, extensively used in the selection of samples of young stellar objects surrounded by infrared-emitting disks of envelopes in star forming regions. \citet{Guarcello13} discuss and combine several of these criteria, together with other indicators of youth such as H$\alpha$ or X-ray emission, to obtain an extensive census of the disk-bearing and accreting population of Cygnus OB2.

Infrared excess-based criteria for the selection of young stellar objects usually rely on their location in the various color-color diagrams that can be defined from the ratios between pairs of fluxes measured in the available photometric bands. Regions of such diagrams are occupied by the different types of young stellar objects generally characterized by the slope of their spectral energy distributions at long wavelengths, allowing for the effect that interstellar extinction has by reddening their intrinsic colors. Since objects devoid of infrared excess display a limited range of intrinsic colors at wavelengths beyond a few microns, which probe the Rayleigh-Jeans regime of the spectral energy distribution, criteria based on color-color diagrams such as those based on Spitzer colors usually work well in characterizing young stellar objects having infrared excesses.

Having access to measurements at shorter wavelengths where infrared excess may be expected to be negligible, we have taken an alternative approach that seeks to simultaneously take into account the intrinsic photospheric emission, the extinction along the line of sight, and the infrared excess of each source. To this end, we express the apparent observed magnitude $m_\lambda$ of a star at wavelength $\lambda$ as

\begin{equation}
m_\lambda = M_\lambda + DM + \alpha_\lambda A_K - E_\lambda  \label{mag_fit}$$
\end{equation}

\noindent where $M_\lambda$ is the absolute magnitude of the star, $DM$ is the distance module ($DM = 10.9$, corresponding to a distance of 1.5~kpc), $\alpha_\lambda = A_\lambda / A_K$ is the ratio between the extinction at $\lambda$ and in the $K_S$ band, and $E_\lambda$ is the excess contributed by circumstellar emission at $\lambda$. We use the \citet{Cardelli89} extinction law with $R_V = 3.1$. The absolute magnitudes used in Eq.~\ref{mag_fit} is obtained from the Pisa evolutionary tracks \citep{Tognelli11}, complemented with the \cite{Baraffe15} evolutionary tracks for masses below $0.2$~M$_\odot$. Absolute magnitudes from the Pisa tracks were derived from the model temperatures and luminosities using the bolometric corrections from the MESA stellar model grids \citep{Paxton11,Paxton13,Paxton15,Choi16,Dotter16}. In their study of young stellar objects with excess in the Spitzer bands, \cite{Beerer10} show that most young stellar objects are roughly clustered along the 1~Myr isochrone in the intrinsic color-magnitude diagram of sources for which spectral classification is available. We therefore adopt that age as well, and use the 1~Myr isochrone when fitting the photometry using Eq.~\ref{mag_fit}, although we must keep in mind that individual sources may be only poorly represented by that isochrone.

To estimate the intrinsic parameters of each star we proceed in general by considering the whole range of stellar masses that define the chosen isochrone. For each mass we compute the extinction $A_V$ that would be required to produce its observed spectral energy distribution at wavelengths expected to be dominated by the photospheric flux, which we obtain by solving Eq.~\ref{mag_fit} as

\begin{equation}
A_K = {{\Sigma_{i=1}^n [\alpha_i (m_i - M_i - DM)] / \sigma_i^2 } \over {\Sigma_{i=1}^n [\alpha_i^2 / \sigma_i^2]}} \label{avfit}
\end{equation}

\noindent where the summation index $i$ refers to all the bands shorter than $K_S$ for which a measurement is available and $\sigma_i$ is the standard error in the magnitude. We discard solutions with $A_K < 0$, and we select the best fit as that corresponding to the mass for which the residual

\begin{equation}
S = \Sigma_{i=1}^n (m_i - M_\lambda - DM - \alpha_\lambda A_K)^2 / \sigma_i^2  \label{residual}$$
\end{equation}

\noindent is minimal.

We have adopted this procedure for all the stars for which we have measurements in at least two bands comprised between $g'$ and $H$, as well as for stars with a measurement in $K_S$ and in another band with shorter wavelength. In this latter case we assume that the $K_S$-band excess is negligible, which may not always be the case.

Some of the stars are undetected in our observations in bands with wavelengths shorter than $K_S$. For those, we have obtained a rough estimate of intrinsic parameters by adopting $A_K = 2$~mag as representative of the range of extinction values found in the region, and we have adopted the mass for which the shorter wavelength available reproduces the magnitude predicted by the isochrone with the adopted extinction.

\begin{figure*}[ht]
\begin{center}
\hspace{-0.5cm}
\includegraphics [width=16cm, angle={0}]{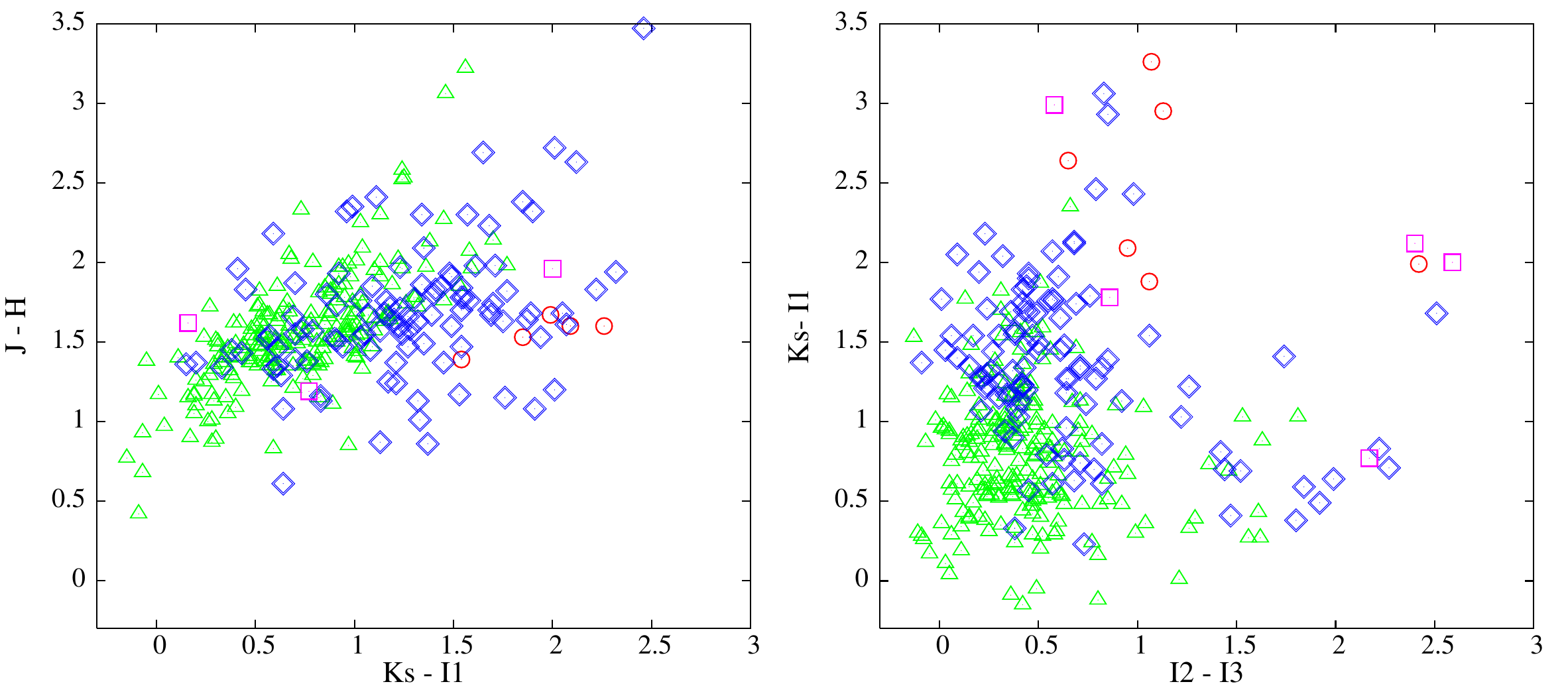}
\caption []{Color-color diagrams of all the YSOs identified in the fields of both DR~18 and ECX~6-21. The two examples shown here are respectively dominated by the effects of extinction $(K_S-I1, J-H)$ and circumstellar excess $(I2-I3, K_S-I1)$. Red circles are Class~I sources, blue diamonds Class~II sources, green triangles transitional sources, and magenta squares Flat Spectrum sources. Sources of the same type tend to cluster on specific regions of each diagram, which nevertheless overlap due to our simultaneous use of all the available photometry to derive the spectral index, $n$, on which our classification is based.}
\label{colcol}
\end{center}
\end{figure*}

Once the intrinsic parameters and the extinction have been obtained with the procedure outlined above, we have estimated the amount of infrared excess from the measurements at all the long wavelength bands starting with $K_S$ if available, or from the band with the shortest wavelength redder than $K_S$ if the latter is not available. We characterize the infrared excess from the usual definition of the spectral index $n = {\rm d} \log \lambda f_\lambda / {\rm d} \log \lambda$, characterizing the infrared excess as an excess of emission over the Rayleigh-Jeans tail of the black body spectral energy distribution ($f_\lambda \propto \lambda^{-4}$) and using the latter to approximately define the zeropoint $C$ of the Vega magnitude scale at long wavelengths. In this way, using $m_\lambda = C -2.5 \log f_\lambda$ we obtain $n$ from a least-square fit to the extinction-corrected magnitudes:

\begin{equation}
m_\lambda - \alpha_\lambda A_K = (n + 3) \log \lambda + C  \label{specind}
\end{equation}

\noindent where we use the available measurements at all bands ($\lambda$) longwards from $H$. This is the same procedure used by \citet{Greene94}, who propose a classification of YSO into four categories according to $n$: Class I ($n > 0.3$), Flat Spectrum ($-0.3 < n \leq 0.3$), Class II ($-1.6 < n \leq -0.3$), and Class III ($n < -1.6$). 
This is also the classification that we adopt in this work, which has also been used in other Spitzer-based studies, most notably the comprehensive work of \citet{Evans09}. We note that other classification schemes based on Spitzer color-color diagrams differ slightly and shift the boundaries between classes. \citet{Megeath04} delimit the region occupied by Class~I YSOs by $I3 - I4 > 1.2$, $I1 - I2 > 0.4$, which would correspond to $n > 0.43$ and $n > -1.35$ in the respective wavelength ranges defined by those IRAC bands, while \citet{Megeath04} define Class~II sources as those having $I3 - I4 > 0.4$, (corresponding to $n > -1.85$), and $I1 - I2 > 0$ ($n > -3$). Our simultaneous use of magnitude measurements at all the available wavelengths to produce the least-square fit to Eqs.~\ref{specind} blurs the separation between the different types of YSOs in the various color-color diagrams, as shown in Figure~\ref{colcol}. We note however that YSO classification on the basis of the position in color-color diagrams often leads to a different YSO classification of the same object depending on the adopted color criterion, simply reflecting the variety of shapes found among the spectral energy distributions of YSOs surrounded by circumstellar disks and envelopes.

In addition to the four categories described above, YSOs with large central cavities devoid of hot dust are classified as transitional disks \citep{Espaillat14}, characterized by a lack of infrared excess at the short wavelengths where the emission of hot dust grains would peak. The spectral energy distributions of transitional YSO display a wide variety of shapes, and the shortest wavelength at which the infrared excess appears also varies depending on the size of the central cavity \citep{Nagel21}. To identify these YSOs in our sample we use a second spectral index $n_2$ defined by the same equation \ref{specind}, but computed from the magnitudes at the two longest wavelengths available, and classify as transitional those YSOs having $n < -1.6$, $n_2 > -1.6$.

The application of the analysis of the spectral energy distributions of the sources detected provide us with a catalog of 349 Class~I, Flat Spectrum, Class~II, and transitional sources in the imaged field around DR~18, and 290 in the field around ECX~6-21. We note that these numbers are the result of a careful visual inspection of each source initially selected as candidate YSO in those categories. We found significant contamination by false detections caused by nebulosity, particularly at the longer wavelengths where the nebulosity is brighter and the diffraction-limited core of the point-spread function is broader, despite the automated removal of extended sources described at the beginning of this Section. We find that visual inspection is critical to avoid the spurious increase of the census of both globules and their nearby nebulosity due to the chance alignment of unrelated sources with local peaks of bright nebulosity at long wavelengths.

\section{The distribution of young stellar objects \label{distribution}}

\begin{figure*}[ht]
\begin{center}
\hspace{-0.5cm}
\includegraphics [width=12cm, angle={0}]{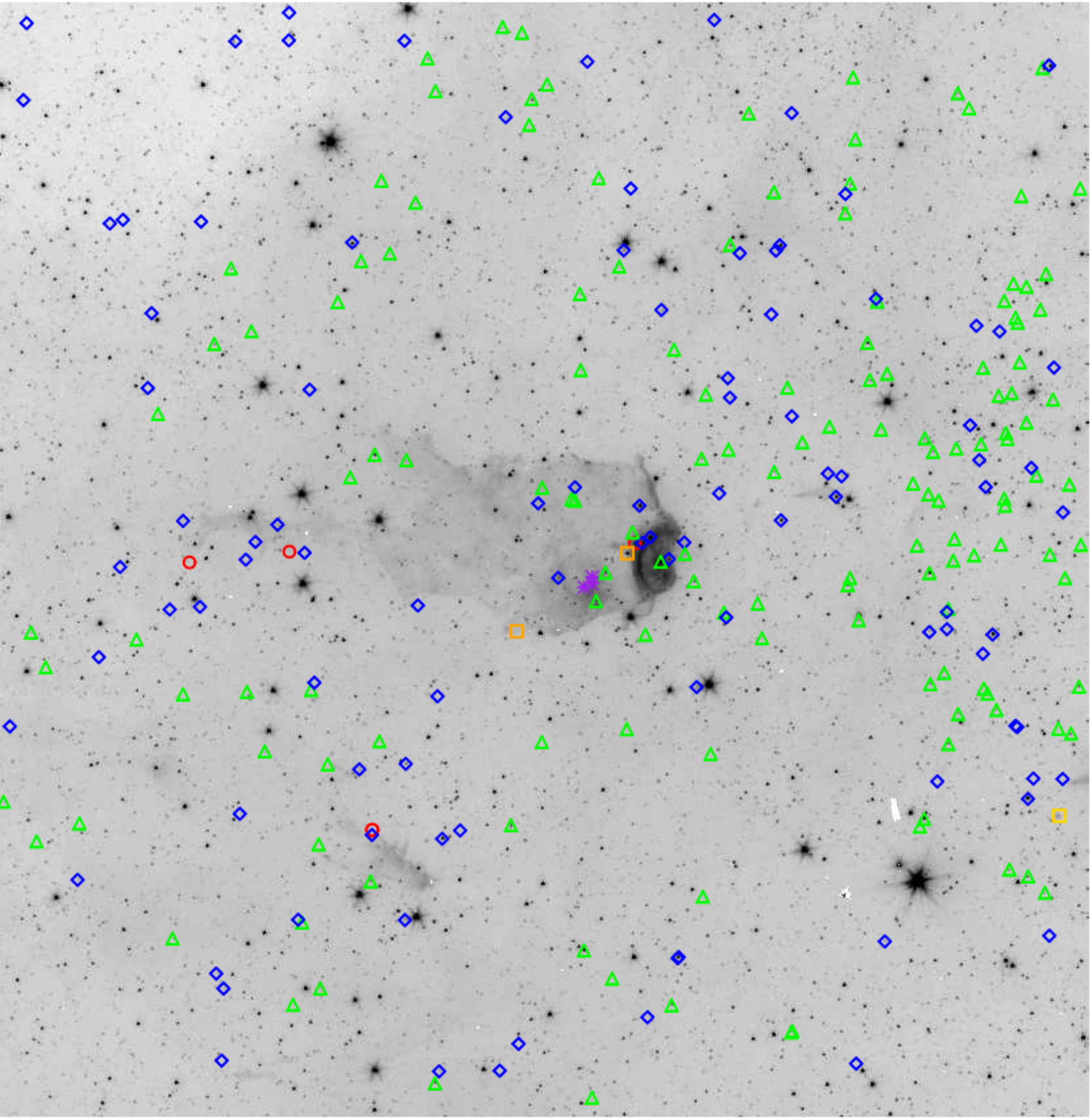}
\caption []{Young stellar objects identified in the field of DR~18. The symbols represent different types of YSOs: Class~I (red circles), Flat Spectrum (amber squares), Class~II (blue diamonds) and transitional (green triangles). The background image is the Spitzer $I2$ IRAC band.}
\label{DR18_YSOs}
\end{center}
\end{figure*}

\begin{figure*}[ht]
\begin{center}
\hspace{-0.5cm}
\includegraphics [width=12cm, angle={0}]{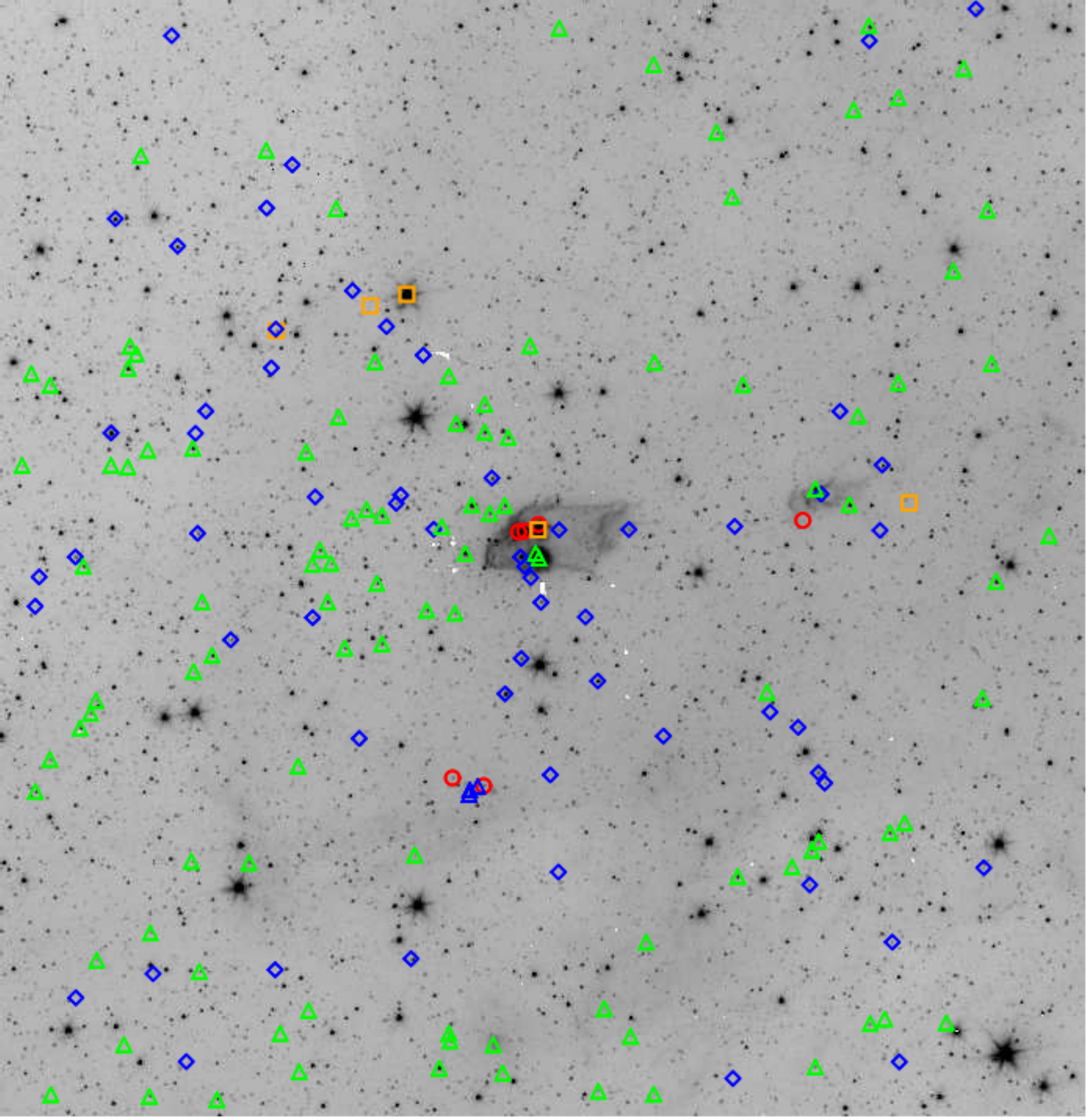}
\caption []{Same as Fig.~\ref{DR18_YSOs}, but now for ECX~6-21.}
\label{ECX6-21_YSOs}
\end{center}
\end{figure*}

Figures~\ref{DR18_YSOs} and \ref{ECX6-21_YSOs} show the location of the various classes of YSOs around each globule. The all-pervading distribution of YSOs reproduces the general traits found by \citet{Guarcello13}, including the higher areal density in the regions of each field closer to the projected center of Cygnus~OB2, this is, west of DR~18 and east of ECX~6-21. The vast majority of members of the distributed population are Class~II and transitional YSOs, whereas the younger Class~I and Flat Spectrum YSOs are present almost exclusively in the globules. We note that there is no obvious excess of Class~II or transitional sources projected on or near DR~18 or ECX~6-21, indicating that most of those that appear projected within the contours of each globule may actually be members of the general Cygnus~OB2 population, not physically related to the globules.

\subsection{YSOs in DR~18 \label{yso_DR18}}

\begin{figure*}[ht]
\begin{center}
\hspace{-0.5cm}
\includegraphics [width=14cm, angle={0}]{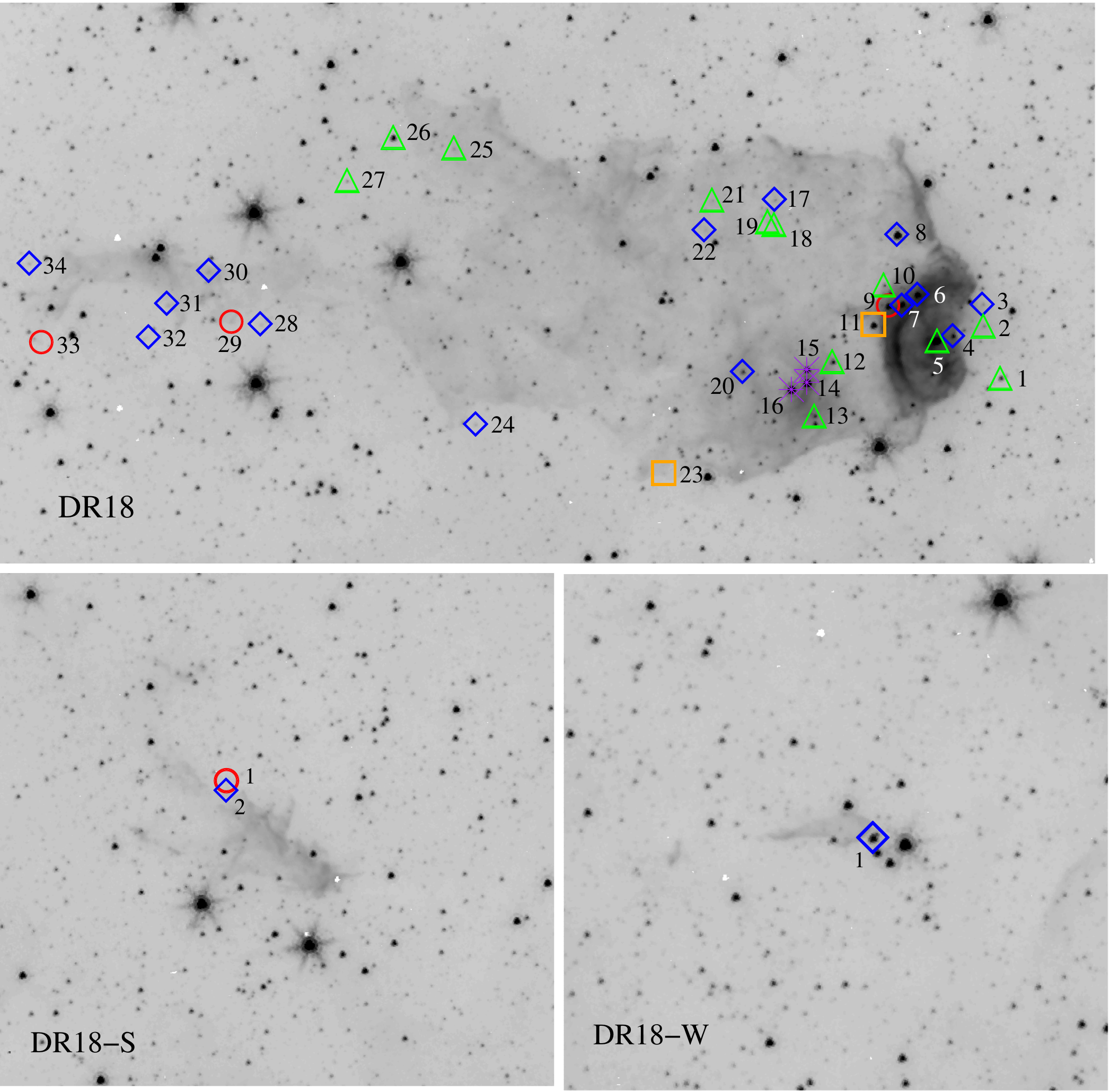}
\caption []{A close-up view of DR~18, DR~18-W and DR~18-S, with their possible associated YSOs. Symbols are as in Fig.~\ref{DR18_YSOs}, with the addition of purple asterisks marking the three Class~III YSOs that we consider as likely members of the stellar population of the globule. The background grey scale image is in the IRAC 4.5~$\mu$m ($I2$) band.}
\label{DR18_mosaic}
\end{center}
\end{figure*}

Table~\ref{YSOs_DR18} lists the YSOs identified within the contours of DR~18 and the small globules DR~18-W and DR~18-S, as well as some objects located just outside their contours but likely associated with them. Their location is shown in Figure~\ref{DR18_mosaic}. Almost all the objects are also plotted on the wider-field image in Figure~\ref{DR18_YSOs}, but we also have added three objects with Class III spectral energy distribution (DR~18-14, DR~18-15, and DR~18-16) that may belong to the globule, as discussed below.

\subsubsection{DR~18 main globule\label{DR18main}}

\begin{figure}[ht]
\begin{center}
\hspace{-0.5cm}
\includegraphics [width=8.5cm, angle={0}]{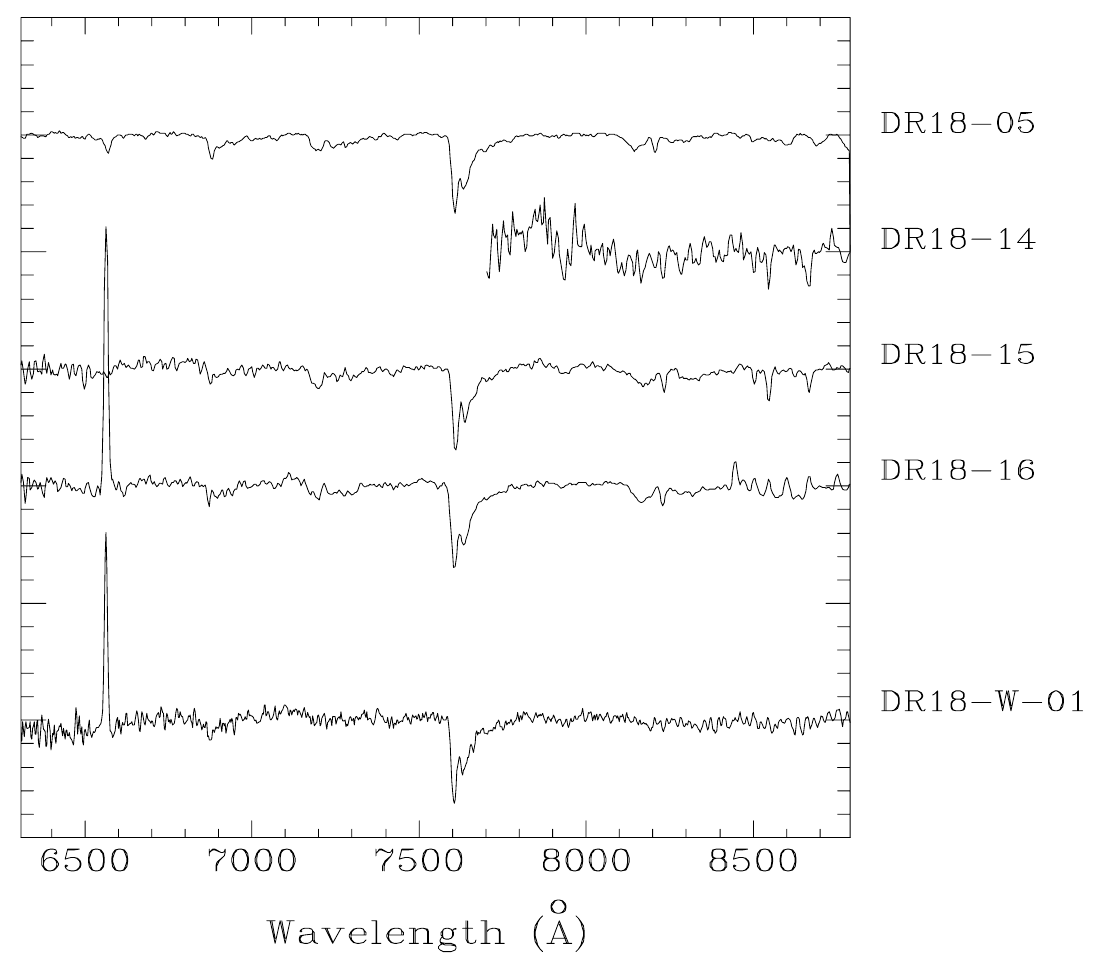}
\caption []{Red visible spectra of DR~18-05 (the early B-type star causing the erosion at the head of the globule) and the three Class~III YSOs DR~18-14, DR~18-15 and DR~18-16. Also shown is the spectrum of DR~18-W-01.}
\label{spectra_DR18stars}
\end{center}
\end{figure}

Most of the YSOs in DR~18 are concentrated near the western edge of the globule. The action of star DR~18-05 on that part of the globule is obvious from the bright arc-shaped nebula, as discussed in Section~\ref{DR18ECX621}. A new visible spectrum of DR~18-05 leads us to classify as a B2 star (Figure~\ref{spectra_DR18stars}), slightly later than the spectral type B0.5$\pm$0.5 assigned in \citet{Comeron99} but consistent within the estimated uncertainty of one spectral subclass. The spectral type implies an effective temperature around $\sim 20,000$~K \citep{Pecaut13} and a mass above the 3.7~M$_\odot$ that we estimate from the best fit to the photometry. Its infrared excess at wavelengths shorter than 8~$\mu$m is mild and it would classify it as a Class~III YSO, but the noticeable increase at 8~$\mu$m placing it in the transitional category. 

The time needed for the B2 star to carve a cavity of the size of the arc-shaped nebula is not in conflict with its youth. Assuming that the star was born near the edge of the cloud and that the freshly ionized gas can escape freely, the cavity reaches a radius $R$ in a time $t = (4 \pi n_H R^3) / (3 S_{UV})$, where $S_{UV}$ is the ionizing flux of the star and $n_H$ the volume density of hydrogen nuclei. Taking $S_{UV} = 5.9 \times 10^{45}$~s$^{-1}$ for a B2 star from the PoWR Grid models \citep{Hainich19}, $n_H = 2 <n_{H_2}> = 4,000$~cm$^{-3}$ as the average density of the DR~18 globule \citep{Schneider16}, $R = 0.22$~pc as the radius of the cavity measured in our images, and neglecting the absorption of ionizing photons by dust, we obtain $t \sim 30,000$~years. This is most certainly a generous lower limit as the adopted density, which is averaged over the total area of the cloud including rather tenuous regions, is unlikely to be representative of the birthplace of DR~18-05, where the density may be expected to have been significantly higher.

\begin{figure}[ht]
\begin{center}
\hspace{-0.5cm}
\includegraphics [width=8.5cm, angle={0}]{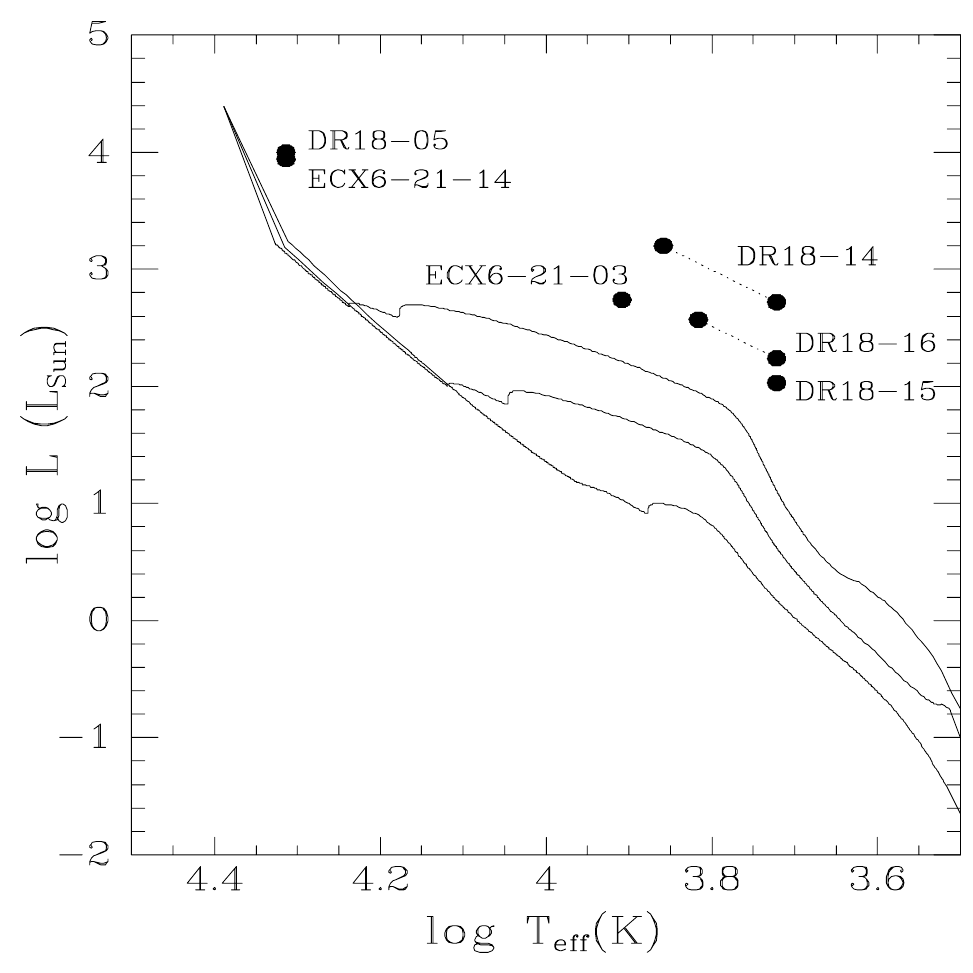}
\caption []{Temperature-luminosity of all the stars spectroscopically classified in the red spectral region in both globules. The Pisa isochrones~\citep{Tognelli11} for the ages 1, 3, and 10~Myr, shifted to a distance modulus $DM = 10.9$ are plotted. Due to the very uncertain types for DR~18-14 and DR~18-16, their positions in the diagram are plotted for the earliest and latest spectral types consistent with the spectra and joined with a dashed line.}
\label{hr}
\end{center}
\end{figure}

\begin{figure*}[ht]
\begin{center}
\hspace{-0.5cm}
\includegraphics [width=12cm, angle={0}]{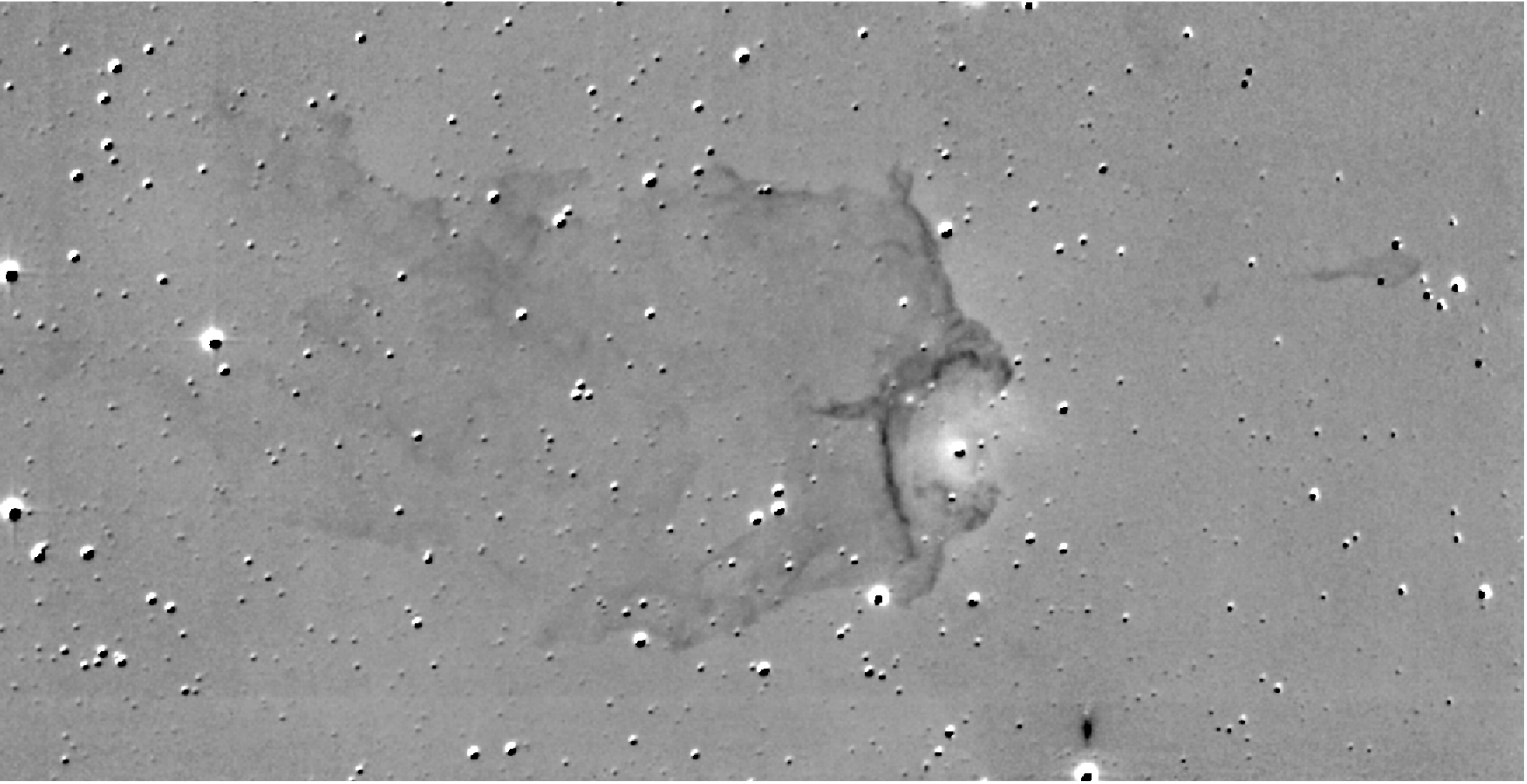}
\caption []{Difference between the images of DR~18 taken through the Br$\gamma$ and the H$_2$ filters, highlighting the regions dominated by ionized hydrogen emission (light) and excited molecular hydrogen emission (dark).}
\label{DR18_Brg-H2}
\end{center}
\end{figure*}

The availability of a spectral classification allows us to use the corresponding temperature and intrinsic colors to estimate extinction, luminosity and age independently of the fit of the spectral energy distribution to a preselected isochrone. With the adopted distance module, the comparison with the evolutionary tracks places it above the 1~Myr isochrone (see Fig.~\ref{hr}), seemingly implying an extremely young age given the short pre-main sequence evolutionary timescale of such massive stars. However, the steep dependency of the temperature on the spectral type makes the precise location of DR~18-05 with respect to the isochrone uncertain. A shift by only half a spectral class from B2 to B1.5, still consistent with our spectrum, increases the temperature by almost 3,000~K \citep{Pecaut13} while leaving the derived luminosity almost unchanged, placing the star on the zero-age main sequence. The higher temperature implies in turn a higher dissociating flux and a faster rate of erosion of the cloud for a given density.  

The clustering of various types of YSOs around DR~18-05 suggests that star formation has been active until the recent past and may still be ongoing. Object DR~18-01, outside the contours of the globule, is highly reddened and most likely background. DR~18-02 and DR~18-03 are less obscured, are projected closer to the contour of DR~18, and they may be low-mass YSOs physically related to it. The characteristics of DR~18-04 and its close projected proximity to DR~18-05 also suggest that it is a roughly solar-mass YSO belonging to the globule population.

The chain of YSOs formed by DR~18-06, DR~18-07, DR~18-09, and DR~18-11 is roughly aligned along the border of the arc nebula. We find very similar masses for DR~18-06, DR~18-07 and DR~18-11. The 1~Myr isochrone yields $M = 3.7$~M$_\odot$ for the three, with DR~18-09 being less massive at 0.7~M$_\odot$. They are also among the most deeply embedded objects in the globule. In addition, the Class~II YSO DR~18-07 and the Flat-spectrum YSO DR~18-11 display very strong Br$\gamma$ emission. Using Eq.~\ref{ew}, we estimate $EW({\rm Br}\gamma) \simeq (135 \pm 16)$~\AA \ for DR~18-07 and $EW({\rm Br}\gamma) \simeq (235 \pm 14)$~\AA \ for DR~18-11. Their characteristics suggest that they are the youngest members of the group to which DR~18-05 belongs, with star formation having proceeding inward into the globule. DR~18-10 may be part of that group, but its relatively blue $JHK_S$ colors imply are consistent with a foreground Class~II YSO belonging to the general Cygnus~OB2 population. Although somewhat disconnected from the rest of the group, the Class~II YSO DR~18-08 and the transitional YSO DR~18-12 also have masses somewhat above 3~M$_\odot$ when assuming an age of 1~Myr. 

The lightly reddened small group formed by DR~18-14, DR~18-15 and DR~18-16 stands out at all wavelengths. None of those stars has strong infrared excess and their spectral energy distributions classify them as Class~III. Their degree of reddening and derived mass are similar when assuming an age of 1~Myr, and they are in turn close to those derived for DR~18-05. However, unlike DR~18-05 this group does not seem to have had a strong impact on its surroundings apart from causing a brightness enhancement to its south and southwest, and a ridge of nebulosity to the west whose brightness pattern points to those stars as the source of illumination. This is consistent with the spectra of those three stars (Fig.\ref{spectra_DR18stars}). DR~18-16 shows strong emission in H$\alpha$ and in all the lines of the Paschen series. The hydrogen emission lines and the lack of other discernible absorption features in its red spectrum, in particular OI absorption at 7774~\AA , suggests that DR~18-16 is an accreting YSO with an underlying spectral type F5 or later. Concerning the other two members of the group, we classify DR~18-15 as early K-type based on the strength of the CaII triplet, absence of OI and Paschen absorption lines, and weakness of the H$\alpha$ absorption. Our spectrum of DR~18-14 is too noisy to allow a more precise classification, but the CaII triplet is clearly detected, indicating a spectral type later than early F. The ridge of nebulosity east of this group contains a much more embedded ($A_K = 2.8$) Class~II source, DR~18-20, which at an age of 1~Myr would have a mass of 5~M$_\odot$. However, the whole group is likely to be significantly younger, as shown from the position of its members in the temperature-luminosity diagram (Fig.~\ref{hr}) derived from the constraints on their spectral types.

Although some low-mass Class~II and transitional YSOs are found within the contours of the globule, their areal density is indistinguishable from that of the same types of objects in the surrounding Cygnus~OB2 population near the position of DR~18, and they are likely members of that distributed population. Lacking any obvious signs of association with nebulosity, we do not consider that they provide solid evidence for other star formation sites in the globule. However, we note the presence of a loose group of very faint, very low mass YSOs near the eastern edge of DR~18. The spectral energy distribution of two of them, DR~18-29 and DR~18-33, places them in Class~I due to their red $K_S - I1$ colors and their clear detection in $I3$, although being close to the detection limit of the observations in the latter band it is difficult to assess whether or not the emission at their position is slightly extended. Among the rest, DR~18-31 and DR~18-32 are doubtful having a very red $K_S - I1$ color, but neutral or even blue $I1 - I2$. Again, the detections in those IRAC bands are near the detection limit and should be taken with caution.

Three of the YSOs that we identify in DR~18 appear in the catalog of protostars of \citet{Kryukova14}. This includes the Class~I YSO DR~18-09 discussed above, but also the B2 star DR~18-05, whose classification in that work is due to the strong infrared excess detected at 24~$\mu$m. The third object in common with our list is DR~18-34, a faint object at the tail of DR~18 with a moderate infrared excess in the IRAC bands that leads to our classification as Class~II. However, its detection at 24~$\mu$m places it among the protostar candidates in \citet{Kryukova14}.

The absence of most of the YSOs detected in the area of DR~18 from the catalog of \citet{Kryukova14} is to be expected due to their moderate excess emission. Conversely, we do not identify as YSO the remaining eight protostellar candidates listed in that work. Examining the Spitzer images at their cataloged positions show that they appear to be local brightenings of nebulosity mimicking the colors of protostars, especially at the longest wavelengths.

The coincidences with our list of YSOs within the area of DR~18 are more numerous when comparing it to the results of \citet{Guarcello13}, as expected given their use of visible and infrared imaging. The seven sources in common again include DR 18~05 and DR~18-09, as well as the Class~II source DR~18-08, the Flat Spectrum source DR~18-11, and the transitional YSOs DR~18-13 and DR~18-26. The seventh YSO is our Class~III YSO DR~18-16, discussed above. As in \citet{Kryukova14}, we also find a number of probably spurious detections in the catalog of \citet{Guarcello13}, highlighting the difficulties of reliable YSO identifications in the presence of structured nebulosity.

\subsubsection{Smaller globules near DR~18 \label{DR18_around}}
 
DR~18-W is \#10 in the proplyd-like category in \citet{Schneider16}. A Class~II YSO, DR~18-W-01, appears projected just outside its head. Our best fit yields a mass of 5.7~M$_\odot$ at 1~Myr. The red-visible spectrum of DR~18-W-01 (Fig.~\ref{spectra_DR18stars}) does not provide any significant further constraints on the nature of DR~18-W-01 besides confirming its H$\alpha$ emission, suspected from the narrow-band photometry, for which we measure an equivalent width of $22 \pm 2$~\AA . The extinction in its direction derived from the fit to the infrared colors is almost three times as high as that of DR~18-05, despite DR~18-W-01 being outside the contour of its globule. This is nevertheless most likely due to significant excess emission in the $K$-band, as suggested by its classification as a Flat Spectrum source by \citet{Guarcello13}, who detect it in all the visible bands down to $u$. DR~18-W-01 is therefore probably a lower-mass YSO with very red infrared colors due to strong infrared excess, although the available spectroscopy does not provide any other significant constraints. We find no other YSOs associated with DR~18-W, and it is therefore possible that
DR~18-W-01 is the only star that has formed in this proplyd-like globule.

DR~18-S has a less defined, but still recognizable head-tail shape oriented similarly to DR~18 and DR~18-W. We find two very faint objects, with best-fit masses near the brown dwarf limit, projected on its tail and very close to each other. DR~18-S-01, also identified by \citet{Kryukova14}, is classified as Class~I mainly due to its faint detection at $5.8$~$\mu$m, whereas the identification of DR~18-S-02 as a Class~II YSO is based to its very red $K_S - I1$ color, the detection at $I2$ being somewhat tentative. There are no morphological indications in the nebulosity of a physical association between the two sources and DR~18-S (which would anyway be subtle at most for such low-mass YSOs), and future, deeper observations providing higher quality flux measurements may question the classification presented here. \citet{Kryukova14} also identify two other Class~I objects within the contours of DR~18-S {and \citet{Guarcello13} add a possible Class~II YSO, their number 1835, none of which we recover. We note that the DR~18-S nebula shows abundant small-scale structure in IRAC images, easily leading to the misidentification of small extended emission peaks as YSOs as discussed in Section \ref{DR18main}.} 

\begin{figure*}[ht]
\begin{center}
\hspace{-0.5cm}
\includegraphics [width=12cm, angle={0}]{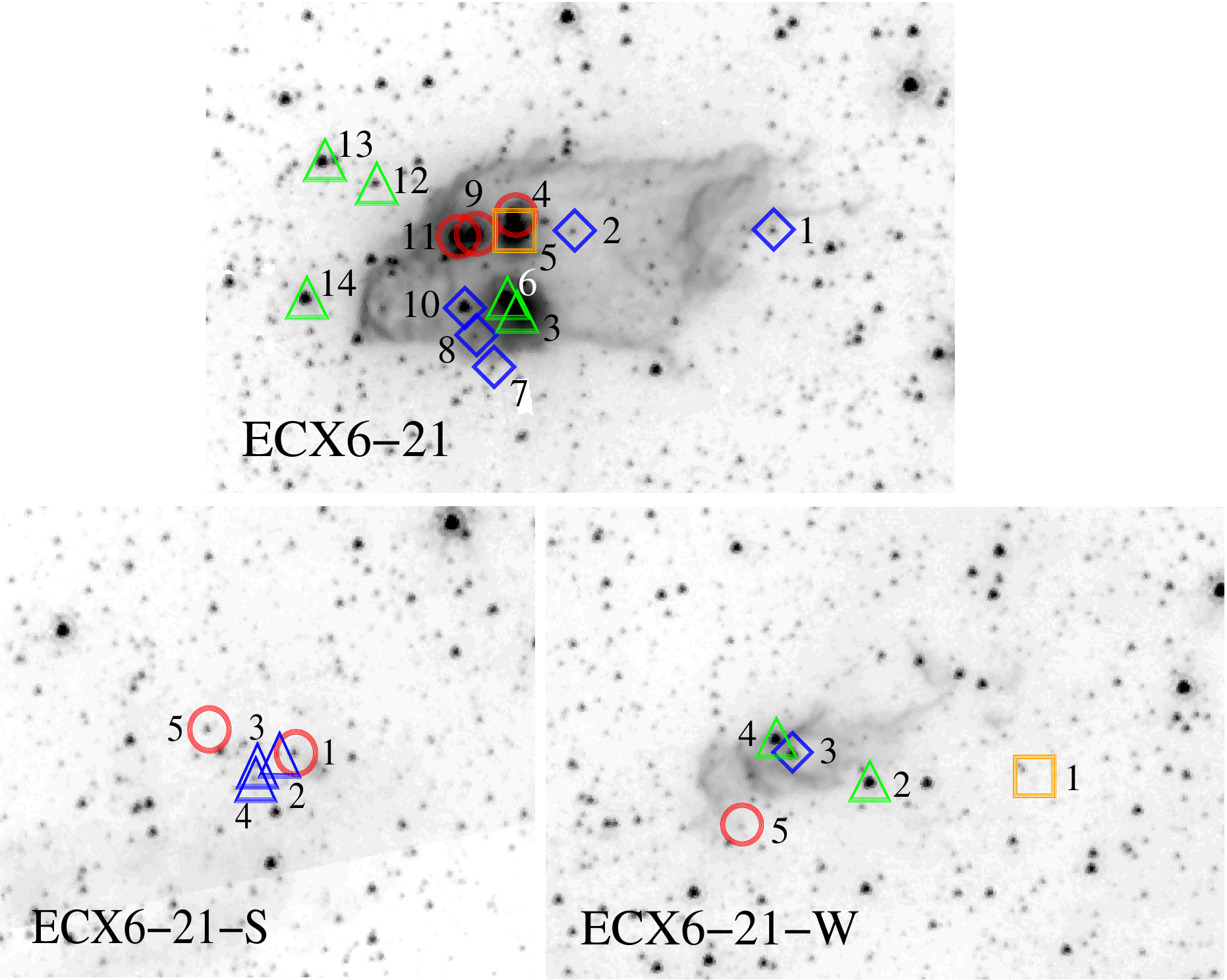}
\caption []{Close-up view of ECX~6-21, ECX~6-21-W, and ECX~6-21-S, with their possible associated YSOs. Symbols are as in Fig.~\ref{DR18_YSOs}. The background grey scale image is in the IRAC 4.5~$\mu$m ($I2$) band.}
\label{ECX6-21_mosaic}
\end{center}
\end{figure*}

\subsection{YSOs in and near ECX~6-21 \label{yso_ECX6-21}}

\subsubsection{ECX~6-21 main globule\label{ECX6-21main}}

\begin{figure}[ht]
\begin{center}
\hspace{-0.5cm}
\includegraphics [width=8.5cm, angle={0}]{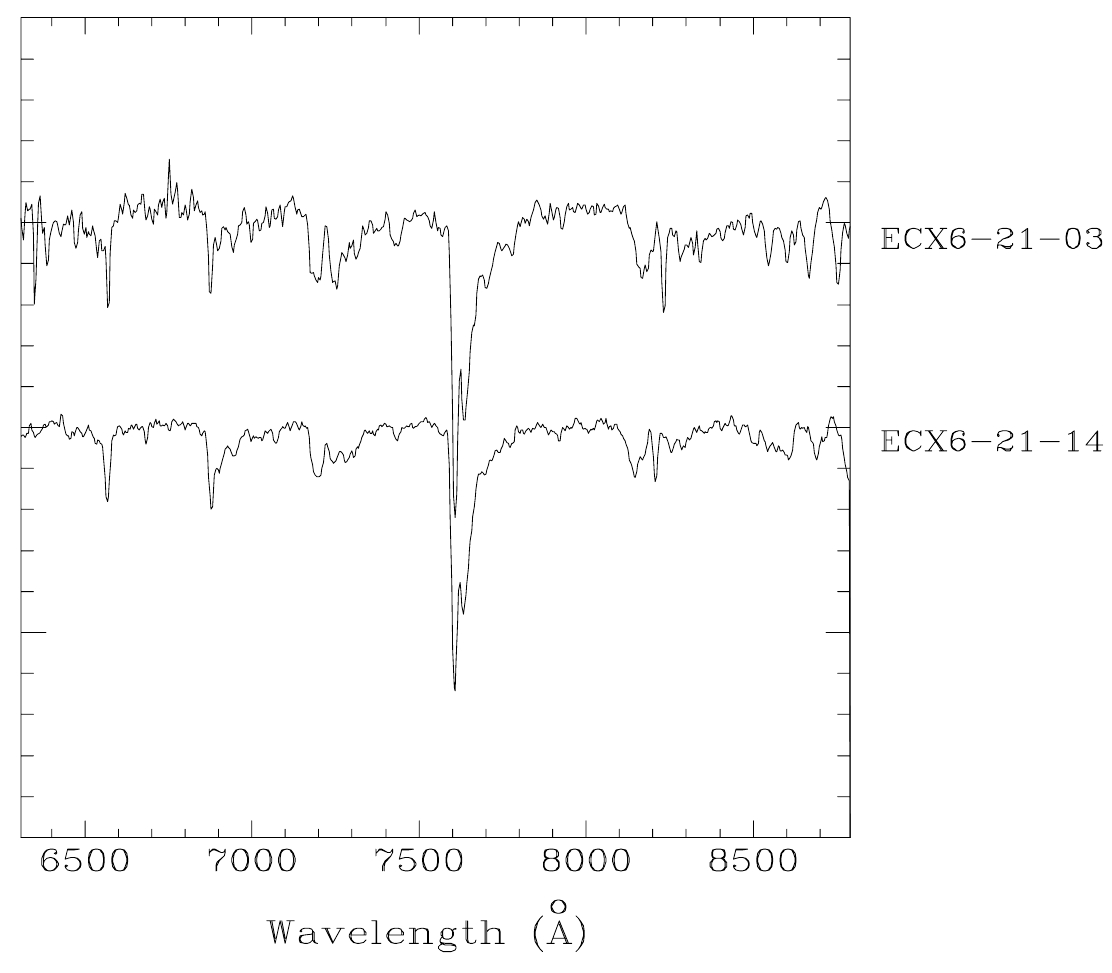}
\caption []{Red-visible spectra of two sources in ECX~6-21. ECX~6-21-03 is a mid A-type YSO in the southern star forming site of the globule. ECX~6-21-14 is an early B-type transitional YSO with a spectrum very similar to that of DR~18-05, located near the position of the peak brightness of the H$\alpha$, Br$\gamma$, and centimeter radiocontinuum that delineates the contour of the globule.}
\label{spectra_ECX6-21stars}
\end{center}
\end{figure}

In their detailed study of ECX~6-21, \citet{Ramachandran17} discuss the properties of the six YSOs that they identify in the globule, almost all of them deeply embedded. Five of them are detected in at least our $K_S$-band image, but ECX~6-21-09 (object \#5 in \citet{Ramachandran17}) is detected in the IRAC bands alone. Only one of the objects, ECX~6-21~03, is lightly embedded ($A_K=0.9$) and appears in visible images as well. Its spectrum (Fig.~\ref{spectra_ECX6-21stars}) corresponds to a mid-to-late A type, consistent with the result of the fit to the spectral energy distribution. Based on the temperature and intrinsic colors derived from its spectral type, the assumed age of 1~Myr is a moderate overestimate (see Fig.~\ref{hr}). Four of the YSOs in the list of \citet{Ramachandran17} are also listed by \citet{Guarcello13}.

In our narrow-band images we detect the H$_2$ knots reported by \citet{Varricatt10}, most likely associated with ECX~6-21-05, which we classify as a flat-spectrum source. We note nevertheless that two additional H$_2$ knots appear to the northeast of the ECX~6-21-05/ECX~6-21-04 pair, as can be seen in Fig.~\ref{ECX6-21_Brg-H2}. Their driving source is uncertain, but based on their proximity we consider it likely that they are associated with one of the Class~I YSOs ECX~6-21-09 or ECX~6-21-11, the latter being the only protostar candidate identified by \citet{Kryukova14} in this globule.

The masses listed in Table~\ref{YSOs_ECX6-21}, estimated as described in Section~\ref{ysosel} using the 1~Myr isochrone, are in the same range as those obtained by \citet{Ramachandran17} using an independent spectral energy distribution fitting method, although both the ages and the amount of foreground extinction that they obtain are much lower. As noted by \citet{Verma03} and \citet{Ramachandran17}, star formation activity in ECX~6-21 appears to be concentrated on two sites, with ECX~6-21-04, 05, 09, and 11 in the northern site and ECX~6-21-03, 06, and 10 in the south. ECX~6-21-06 is not included in the list of \citet{Ramachandran17}, but its spectral energy distribution and derived properties are very similar to those of ECX~6-21-03, both being transitional sources in our classification. The difference in the type of YSOs (Class~I and Flat spectrum in the northern site, Class~II and transitional in the southern one), degree of obscuration (with the southern sources ECX~6-21-03 and ECX~6-21-06 being detectable at visible wavelengths), and presence of discrete H$_2$ knots only in the northern group are all indications that the southern group is older than the northern one. This is also found by \citet{Ramachandran17}, who find extremely young ages, possibly even below 10,000~years, for the stars in the northern group. Furthermore, the three YSOs in the northern group that have $K_S$-band detections seem to be intensely accreting as inferred from their $[{\rm Br}\gamma] - [{\rm H}_2]$, with estimated equivalent widths $EW({\rm Br}_\gamma) \sim 70$~\AA , $110$~\AA , and 300~\AA \ for ECX~6-21-04, 05, and 11, respectively.

We identify four additional Class~II sources, all of them with low masses, projected within or near the contours of ECX~6-21. ECX~6-21-01 is located on the tail of the globule just outside its edge, but the derived extinction, $A_K = 2.3$, makes a physical association with the globule unlikely. ECX~6-21-02 is also heavily obscured as inferred from its non detection at wavelengths shorter than 2~$\mu$m, but its position near the geometric center of the globule make membership more likely. Membership is also uncertain for ECX~6-21-07 and 08, near the southern star forming site. The light extinction toward ECX~6-21-07 and the higher one toward ECX~6-21-08, which is just inside, are as expected given their respective positions just outside and inside the contour of the globule. No other low-mass objects, either suspected or likely members of the globule population, are detected.

We finally note the presence of ECX~6-21-12, 13 and 14, three intermediate- or solar-mass transitional YSOs outside the eastern edge of the globule. We place all three in the category of  transition disks with moderate excesses similar to those of Class~II, which is the classification assigned by \citet{Guarcello13} to the first two, whereas ECX~6-21-14 is classified as flat spectrum / transition disk in that work. ECX~6-21-12 and 14 have light extinction that makes them detectable in the visible, while ECX~6-21-13 is more reddened and most likely background. The visible spectrum of ECX~6-21-14 (Fig.~\ref{spectra_ECX6-21stars}) and its spectral energy distribution are very similar to those of DR~18-05. Its classification as a B2 star places it near the zero-age main sequence at the adopted distance of the region. Its position in the sky near the peak brightness of the crescent of H$\alpha$ and Br$\gamma$ emission that limits the eastern edge of ECX~6-21 suggests that ECX~6-21-14 is physically associated with the globule. Its effects on the molecular gas are nevertheless far less dramatic, if at all existing, than those caused by DR~18-05 on its parental globule, leading us to suspect that ECX~6-21-14 is located at some distance behind the globule as seen from our vantage point.

\subsubsection{Smaller globules near ECX~6-21 \label{ECX6-21_around}}

The globule ECX~6-21-W is much smaller but shares the same orientation as ECX~6-21. Somewhat paradoxically, the two presumably least evolved YSOs in its immediate projected vicinity, the Class~I ECX~6-21-W-05 and the Flat-spectrum ECX~6-21-W-01, appear outside its contours. Both are faint but clearly detected as point sources at 8~$\mu$m, their foreground extinction is low, and their masses may be substellar if their age is significantly younger than the adopted 1~Myr. The transitional sources ECX~6-21-W-02 and 04 are located near the borders of the globule and both are strong Br$\gamma$ emitters from the difference in their narrow-band magnitudes. We estimate $EW({\rm Br}\gamma) = 140$~\AA \ for ECX~6-21-W-02 and $110$~\AA \ for ECX~6-21-W-04, and both seem to be intermediate-mass YSOs with masses above solar.

\begin{figure*}[ht]
\begin{center}
\hspace{-0.5cm}
\includegraphics [width=12cm, angle={0}]{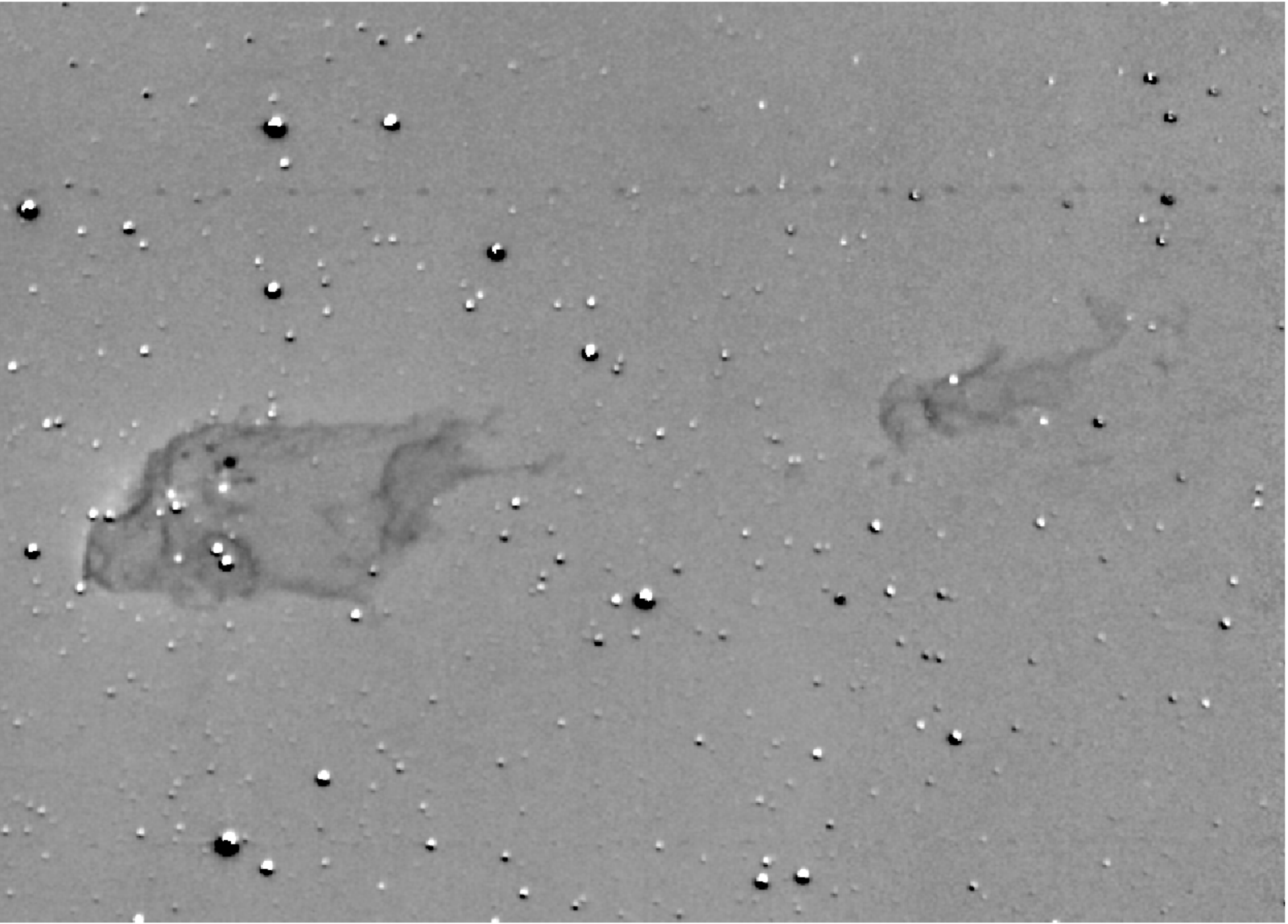}
\caption []{Same as Fig.~\ref{DR18_Brg-H2}, now for ECX~6-21 and ECX~6-21-W.}
\label{ECX6-21_Brg-H2}
\end{center}
\end{figure*}

The structure of ECX~6-21-W, most clearly seen in Figs.~\ref{ECX6-21_mosaic} and \ref{ECX6-21_Brg-H2}, shows a sequence of three arc-shaped rims outlined by H$_2$ emission, all of them with their apses pointing eastward. The easternmost arc is bound by a faint halo in Br$\gamma$, like DR~18 and ECX~6-21, which may be due to its external ionization by the O stars of Cygnus~OB2. Its H$_2$ emission may be at least partly caused by fluorescence of the PDR at the edge of the globule. This explanation does not seem to apply however to the other two arcs, which do not display any detectable Br$\gamma$ halos. However, their origin (and possibly also the origin of the easternmost arc) might be explained by the existence of ECX~6-21-03, a heavily embedded ($A_K = 5.4$) YSO at the focus of the third arc in the east-west direction. We note in addition that the second arc is punctuated by two unresolved knots of H$_2$ emission, suggesting that its H$_2$ emission is caused by shocks. Its position relative to ECX~6-21-W-03 hints at an outflow driven by this YSO as its cause, whereas the third arc may be the working surface of another shock driven by the same source. Interestingly, particle acceleration in reverse shocks is known to give rise to the synchrotron emission detected toward some Herbig-Haro objects \citep{Rodriguez19}, which may provide an explanation to the conflicting classifications of ECX~6-21-W as a conventional HII region or a non-thermal radiosource depending on the diagnostic technique employed, as we noted in Section~\ref{DR18ECX621}. If ECX~6-21-W-03 is indeed the driving source of an outflow, our spectral energy distribution placing it in Class~II would be somewhat surprising as it suggests a more evolved YSO than the typical jet driving source. \citet{Kryukova14} classify it as a protostellar candidate, with a very strong excess at 24~$\mu$m which may nevertheless be contaminated by the bright and compact emission of the globule. On the other hand, the lack of heavily disrupting effects on the environment that would be expected by a star of 3.3~M$_\odot$, which we obtain from the best fit to the spectral energy distribution assuming an age of 1~Myr, hints to a much lower actual mass and a much younger age, more consistent with the typical ages estimated for jet-driving YSOs. We note that the limited amount of measurements for this source (which is undetected below the $H$ band and appears saturated at 8~$\mu$m) makes both its mass estimate and its intrinsic spectral energy distribution particularly uncertain, not ruling out the possibility that it could actually be a low-mass Class~I YSO.

The region that we call ECX~6-21-S is extended and diffuse. The density of YSOs projected on it is not noticeably higher than elsewhere in the field, and most of them belong almost certainly to the general Cygnus~OB2 population. However, we notice the existence of a small, tight group of five faint sources, belonging to classes~I and II, in a region centered near $\alpha(2000) = 20^h 34^m 30^s$, $\delta(2000) = +41^\circ 11' 05''$ and otherwise almost devoid of nebulosity. The classification of the sources ECX~6-21-S-01, 02 and 05 is somewhat tentative, as they are detected only at $K_S$ and longer wavelengths and, according to the procedure described in Section~\ref{ysosel}, we have assumed $A_K = 2.0$~mag for them, rather than deriving $A_K$ from the fit. The extinction may be an overestimate in a region that shows faint nebulosity only, and a fit using a lower value of $A_K$ would yield even lower masses and stronger infrared excesses. This most certainly applies at least to ECX~6-21-S-01, which is among the protostellar candidates of \citet{Kryukova14} and is also noted as Class~I by \citet{Guarcello13}. The detection of the brightest source of the group, ECX~6-21-S-04, in the $H$ band with $(H-K_S) =1.38$, confirms an apparently high extinction. We note that \citet{Guarcello13} classify it as Class~I. The same classification is assigned by them to the fifth YSO in the group, ECX~6-21-05, with which we agree. Interestingly, \citet{Kryukova14} added three more sources to this group, which are very faint and with Class~I spectra, and two of them are also included in the catalog of \citet{Guarcello13} and classified as Class~I and Flat Spectrum, respectively. We have included in our list of possible members the object ECX~6-21-S-03, which is detected in all the bands from $g'$ to 4.5~$\mu$m, indicating a level of obscuration that is significantly lower than the any other object in the region. This makes it a very likely foreground, unrelated object. However, the clustering of at least seven faint sources, most of them with significant infrared excess, in a small area otherwise devoid of other signposts of recent star formation probably deserves further attention.

\section{Discussion and conclusions\label{discussion}}

Both DR~18 and ECX~6-21 are active sites of very recent, and possibly ongoing star formation as shown by the presence of YSOs in their earliest stages of evolution. The observations that we have presented here include both globules, as well as other smaller globules in their neighborhood and an abundant sampling of the field around them, which is dominated by the general population of YSOs belonging to the Cygnus~OB2 association that has been the subject of previous studies. Some members of this YSO population had been identified in previous surveys of the region.

As noted in Section~\ref{ysosel}, we have paid special attention to avoiding the inclusion of false detections among our census of YSOs caused by small-scale peaks of local nebulosity, which would contaminate preferably the Class~I and Flat Spectrum populations. We found this to affect, to some extent, the YSO census produced in previous large-scale surveys of the Cygnus~OB2 and Cygnus~X region in the areas studied here. Although a few doubtful cases remain, noted in Sections~\ref{yso_DR18} and \ref{yso_ECX6-21}, nearly all the YSOs that we classify in those categories in this work are confirmed as point sources by visual inspection, especially those found within the contours of the globules.

Both globules contain Class~I and Flat Spectrum YSOs, which are nearly absent from the field population around them. We have discussed the evidence for the association between some Class~II and transitional YSOs with both globules, the most obvious cases being the transitional YSO DR~18-05 that is eroding the head of the DR~18 globule, the small concentration of Class~II and transitional YSOs in the southern star forming site of ECX~6-21, and the two YSOs found within the contour of ECX~6-21-W. Otherwise, we do not identify a significant excess of such objects in the direction of either globule or their immediate surroundings and many, if not most of the Class~II and transitional found within or near their contours may belong to the Cygnus~OB2 population with no physical relation with the globules.  

The temperature-luminosity diagrams of the stars for which spectroscopic classifications are available, indirect arguments such as the rate of erosion of the DR~18 globule by DR~18-05, and the presence of extreme youth manifestations such as outflows in ECX~6-21 and possibly also in ECX~6-21-W, suggest that star formation in both globules started significantly less than 1~Myr ago. Based on the (Class~I + Flat Spectrum) / (Class~II + transitional) ratios the southern star forming site in ECX~6-21 would be oldest, followed by DR~18, and ending with the northern site of ECX~6-21. On the other hand, the small group formed by DR~18-14, DR~18-15 and DR~18-16 has the surprising property of having its three members devoid of near-infrared excess while their locations in the color-temperature diagram indicates an age younger than 1~Myr, less than the typical timescale for disk dissipation \citep{Cloutier14,Wijnen17,Arun19}, although at least one of them, DR~18-16, is still actively accreting. 

The detachment of both globules from the general complex of Cygnus~X and the youth of the YSOs indicates that their early evolution, and perhaps also the stages previous to their formation, have taken place after the globule became externally irradiated. Radiation-driven implosion \citep{Bertoldi89} has been invoked as the trigger of star formation in other globules \citep{Bowler09,Beltran09,Sicilia14,Kun17,Reiter20}, and both DR~18 and ECX~6-21 are good candidates given the conditions in their environments. This might be reflected in the characteristics of their stellar populations. The very young ages of the YSOs embedded in the globules make the determination of their masses very uncertain, but most of the likeliest members of the populations at the heads of the globules, where star formation concentrates, have masses of around one to a few solar masses. No enhancement of the areal density of lower-mass YSOs is detected in the star forming sites of both globules, even though they would be detectable at the sensitivity levels of the observations presented here. This adds to the hints found in other irradiated environments where the formation of low-mass stars seems to be inhibited \citep[e.g.,][]{Comeron07,Reiter20}. Numerical simulations of the effect of irradiation by O and B stars on an initially turbulent magnetized cloud \citep{Arthur11} show that B stars tend to decrease small-scale turbulence and create thick and relatively smooth PDRs, which may favor the formation of higher-mass stars through fragmentation. The dependency on the incident spectrum may explain why other irradiated globules, like IC 1396N \citep{Getman07,Beltran09,Choudhury10} do not show evidences of a top-heavy mass function of their embedded population. 

The YSO populations of DR~18 and ECX~6-21 that we have presented here, as well as our previous study of IRAS~20319+3958 \citep{Djupvik17}, provide a sampling of different evolutionary stages, masses, and circumstellar environments of YSOs, having in common their formation and early evolution embedded in globules externally irradiated by the ultraviolet radiation of the O stars of Cygnus~OB2. The photometric derivation of parameters that we have presented here must be taken as preliminary only, as suggested by the follow-up spectroscopy of selected members. Nevertheless, the sample provides a promising basis for deeper surveys, infrared spectroscopic investigations of the entire embedded population, and modeling of the interaction of the most massive YSOs with their surrounding interstellar medium, aiming to place such populations in the broader context of star formation in different environments.

\begin{acknowledgements}

We are very pleased to acknowledge the excellent support provided by the staff at the Calar Alto Observatory and at the Teide Observatory. Special thanks go to the staff of the Telescopio Nazionale Galileo for obtaining our observations despite the difficult circumstances that the island of La Palma was undergoing due to the eruption of the Cumbre Vieja volcano only a few kilometers away. Constructive comments by an anonymous referee contributed to various improvements in the presentation of our results. FC acknowledges the support and hospitality of the Faculty of the European Space Astronomy Centre (ESAC). NS acknowledges support by the Agence National de Recherche (ANR/France) and the Deutsche Forschungsgemeinschaft (DFG/Germany) through the project "GENESIS" (ANR-16-CE92-0035-01/DFG1591/2-1), as well as support for travel to Calar Alto by the DFG, grant Nr. 394072042. This work is based in part on observations made with the Spitzer Space Telescope, which was operated by the Jet Propulsion Laboratory, California Institute of Technology under a contract with NASA. The Two Micron All Sky Survey (2MASS) is a joint project of the University of Massachusetts and the Infrared Processing and Analysis Center/California Institute of Technology, funded by NASA and the National Science Foundation. This research has made use of the SIMBAD database, operated at CDS, Strasbourg, France.

\end{acknowledgements}

\bibliographystyle{aa} 
\bibliography{DR18_ECX6-21_cit}

\begin{appendix}
\section{Young stellar objects in DR~18 and ECX~6-21}

\begin{landscape}
\begin{table}
\caption{\label{YSOs_DR18}Young stellar objects in DR~18}
\begin{tabular}{lccccccccccccccccc}
\hline\hline
\noalign{\smallskip}
Object No. & $\alpha (2000)$ & $\delta(2000)$ & Class & $g'$ & $r'$ & $i'$ & $J$ & $H$ & $K_S$ & $I1$ & $I2$ & $I3$ & $I4$ & $M ({\rm M}_\odot)$ & $L$ (L$_\odot$) & $A_K$ & $n$ \\
\noalign{\smallskip}\hline\noalign{\smallskip}
\noalign{\medskip \centerline{DR~18} \medskip}
DR~18-01	& 20:35:03.03 & 41:12:44.5 & Trans & -99.9 & -99.9 & -99.9 & 16.96 & 14.66 & 13.38 & 12.25 & 11.81 & 11.39 & 10.78 & 4.4 &  2.7 & 2.76 & -2.5 \\  
DR~18-02	& 20:35:04.10 & 41:13:22.0 & Trans & -99.9 & -99.9 & -99.9 & 17.31 & 15.85 & 15.13 & 14.29 & 13.74 & 13.19 & -99.9 & 0.2 & -0.6 & 0.85 & -1.8 \\  
DR~18-03	& 20:35:04.17 & 41:13:37.5 & II    & -99.9 & -99.9 & -99.9 & 17.86 & 16.39 & 15.65 & 14.38 & 14.06 & 13.41 & -99.9 & 0.2 & -0.6 & 1.16 & -1.6 \\  
DR~18-04	& 20:35:06.05 & 41:13:14.9 & II    & -99.9 & -99.9 & 19.66 & 15.32 & 13.52 & 12.82 & 11.42 & 11.27 & -99.9 & -99.9 & 1.5 &  0.6 & 1.06 & -1.5 \\  
DR~18-05	& 20:35:07.04 & 41:13:11.4 & Trans & 17.78 & 15.06 & 13.52 & 10.47 &  9.70 &  9.20 &  9.35 &  8.92 &  8.50 &  7.25 & 3.7 &  1.9 & 0.53 & -2.2 \\  
DR~18-06	& 20:35:08.30 & 41:13:44.7 & II    & -99.9 & -99.9 & -99.9 & 19.67 & 15.88 & 13.63 & 10.70 &  9.79 &  8.94 & -99.9 & 3.7 &  1.9 & 4.02 & -0.7 \\  
DR~18-07	& 20:35:09.29 & 41:13:36.9 & II    & -99.9 & -99.9 & -99.9 & 19.01 & 15.15 & 13.08 & 10.96 & 10.06 &  9.38 & -99.9 & 3.7 &  1.9 & 3.60 & -1.5 \\  
DR~18-08	& 20:35:09.60 & 41:14:27.8 & II    & -99.9 & -99.9 & -99.9 & 15.28 & 12.59 & 11.22 &  9.57 &  8.68 &  8.07 &  7.10 & 3.7 &  1.9 & 2.06 & -1.0 \\  
DR~18-09	& 20:35:10.16 & 41:13:36.6 & I     & -99.9 & -99.9 & -99.9 & -99.9 & -99.9 & 15.14 & 11.00 &  9.50 &  8.53 & -99.9 & 0.3 & -0.15 & 2.00 &  2.4 \\  
DR~18-10	& 20:35:10.45 & 41:13:51.6 & Trans & -99.9 & -99.9 & -99.9 & 16.94 & 15.20 & 14.35 & 13.31 & 12.59 & -99.9 & -99.9 & 0.3 & -0.3 & 0.95 & -1.6 \\  
DR~18-11	& 20:35:11.09 & 41:13:23.0 & FS    & -99.9 & -99.9 & -99.9 & -99.9 & 18.05 & 14.81 & 10.86 &  9.70 &  9.00 & -99.9 & 3.7 &  1.9 & 5.12 & -0.1 \\  
DR~18-12	& 20:35:13.72 & 41:12:56.5 & Trans & -99.9 & -99.9 & -99.9 & 16.87 & 14.35 & 13.07 & 11.83 & 11.30 & 10.84 & -99.9 & 4.1 &  2.5 & 2.97 & -2.6 \\  
DR~18-13	& 20:35:14.87 & 41:12:17.4 & Trans & -99.9 & -99.9 & -99.9 & 18.53 & 15.47 & 13.99 & 12.53 & 12.02 & 11.33 & -99.9 & 3.3 &  1.5 & 3.27 & -2.5 \\  
DR~18-14	& 20:35:15.34 & 41:12:41.4 & III   & -99.9 & 20.75 & 18.05 & 11.70 &  9.55 &  8.61 & -99.9 &  8.48 &  8.13 &  7.91 & 3.7 &  1.9 & 0.80 & -3.0 \\  
DR~18-15	& 20:35:15.34 & 41:12:50.7 & III   & 21.33 & 17.78 & 15.85 & 11.75 & 10.32 &  9.84 &  9.67 &  9.69 &  9.65 &  9.53 & 3.7 &  1.9 & 0.86 & -3.2 \\  
DR~18-16	& 20:35:16.32 & 41:12:36.3 & III   & 19.43 & 16.42 & 14.68 & 10.90 &  9.83 &  9.12 &  9.18 &  8.62 &  8.17 &  7.78 & 3.7 &  1.9 & 0.66 & -2.4 \\  
DR~18-17	& 20:35:17.40 & 41:14:53.0 & II    & -99.9 & -99.9 & -99.9 & 17.68 & 16.08 & 15.38 & 14.16 & 13.89 & 12.63 & -99.9 & 0.2 & -0.6 & 1.01 & -1.2 \\  
DR~18-18	& 20:35:17.43 & 41:14:35.0 & Trans & -99.9 & -99.9 & -99.9 & 19.38 & 17.05 & 16.01 & 15.28 & 14.95 & 13.59 & -99.9 & 0.2 & -0.6 & 1.70 & -2.2 \\  
DR~18-19	& 20:35:17.84 & 41:14:36.9 & Trans & -99.9 & -99.9 & -99.9 & 18.16 & 16.11 & 15.18 & 14.51 & 14.34 & 13.39 & -99.9 & 0.2 & -0.5 & 1.27 & -2.3 \\  
DR~18-20	& 20:35:19.44 & 41:12:49.3 & II    & -99.9 & -99.9 & -99.9 & 17.26 & 14.96 & 13.99 & 12.65 & 12.00 & 11.57 & 10.22 & 4.6 &  2.7 & 2.78 & -1.6 \\  
DR~18-21	& 20:35:21.38 & 41:14:52.6 & Trans & -99.9 & -99.9 & -99.9 & 18.12 & 16.32 & 15.56 & 14.68 & 14.72 & 13.09 & -99.9 & 0.2 & -0.6 & 1.20 & -1.8 \\  
DR~18-22	& 20:35:21.88 & 41:14:31.1 & II    & -99.9 & -99.9 & -99.9 & 19.77 & 17.54 & 16.67 & 14.99 & 15.14 & 12.63 & -99.9 & 0.2 & -0.6 & 2.00 & -0.6 \\  
DR~18-23	& 20:35:24.44 & 41:11:36.3 & FS    & -99.9 & -99.9 & -99.9 & 19.72 & 17.76 & 17.00 & 15.00 & 15.23 & 12.64 & -99.9 & 0.1 & -1.2 & 1.24 &  0.2 \\  
DR~18-24	& 20:35:36.43 & 41:12:11.5 & II    & -88.8 & -88.8 & -88.8 & -99.9 & 17.36 & 16.17 & 15.11 & 14.73 & -99.9 & 12.27 & 0.2 & -0.6 & 1.99 & -1.3 \\  
DR~18-25	& 20:35:37.83 & 41:15:30.0 & Trans & -88.8 & -88.8 & -88.8 & 17.58 & 15.79 & 15.06 & 14.03 & 14.39 & 12.58 & -99.9 & 0.2 & -0.6 & 0.92 & -1.5 \\  
DR~18-26	& 20:35:41.69 & 41:15:37.5 & Trans & -88.8 & -88.8 & -88.8 & 15.26 & 13.30 & 12.29 & 11.05 & 10.64 & 10.25 &  9.62 & 4.1 &  2.4 & 2.29 & -2.2 \\  
DR~18-27	& 20:35:44.63 & 41:15:06.5 & Trans & -88.8 & -88.8 & -88.8 & 17.71 & 15.87 & 15.11 & 14.26 & 13.97 & 13.85 & 13.29 & 0.2 & -0.6 & 0.98 & -2.2 \\  
DR~18-28	& 20:35:50.15 & 41:13:23.3 & II    & -88.8 & -88.8 & -88.8 & 19.63 & 17.28 & 16.23 & 15.24 & 15.21 & -99.9 & 11.76 & 0.2 & -0.6 & 1.83 & -0.6 \\  
DR~18-29	& 20:35:51.99 & 41:13:24.8 & I     & -88.8 & -88.8 & -88.8 & 19.29 & 17.69 & 17.03 & 14.77 & 15.12 & 12.07 & -99.9 & 0.1 & -1.2 & 1.15 &  0.9 \\  
DR~18-30	& 20:35:53.44 & 41:14:01.6 & II    & -88.8 & -88.8 & -88.8 & 16.74 & 15.58 & 15.26 & 14.43 & 14.52 & 12.30 & -99.9 & 0.2 & -0.6 & 0.61 & -1.0 \\  
DR~18-31	& 20:35:56.11 & 41:13:38.0 & II    & -88.8 & -88.8 & -88.8 & 19.72 & 17.78 & 17.05 & 14.73 & 15.11 & -99.9 & -99.9 & 0.1 & -1.2 & 1.26 & -0.5 \\  
DR~18-32	& 20:35:57.27 & 41:13:13.8 & II    & -88.8 & -88.8 & -88.8 & 19.53 & 17.85 & 17.18 & 15.29 & 15.45 & -99.9 & -99.9 & 0.1 & -1.2 & 1.26 & -1.1 \\  
DR~18-33	& 20:36:04.09 & 41:13:09.9 & I     & -88.8 & -88.8 & -88.8 & 19.79 & 18.26 & 17.50 & 15.65 & -99.9 & 13.11 & -99.9 & 0.07 & -1.4 & 1.15 &  0.5 \\  
DR~18-34	& 20:36:04.87 & 41:14:06.6 & II    & -88.8 & -88.8 & -88.8 & 18.51 & 16.71 & 15.85 & 14.99 & 14.58 & 13.76 & 12.12 & 0.2 & -0.6 & 1.37 & -1.2 \\  
\noalign{\medskip \centerline{DR~18-W} \medskip}
DR~18-W 01	& 20:34:45.81 & 41:14:39.7 & II	& -99.9 & 20.74 & 18.94 & 14.00 & 12.37 & 11.65 & 10.33 &  9.76 &  9.39 &  8.97 & 5.8 & 2.9 & 1.42 & -1.6  \\   
\noalign{\medskip \centerline{DR~18-S} \medskip}
DR~18-S 01	& 20:35:41.93 & 41:07:04.8 & I 	& -88.8 & -88.8 & -88.8 & 20.00 & 18.33 & 17.70 & 15.71 & 15.68 & 13.26 & -99.9 & 0.08 & -1.3 & 1.33 &  0.4 \\   
DR~18-S 02	& 20:35:41.96 & 41:06:57.9 & II	& -88.8 & -88.8 & -88.8 & -99.9 &  18.51 & 17.54 & 15.70 & 15.87 & -99.9 & -99.9 & 0.08 & -1.4 & 1.37 & -1.2 \\
\noalign{\medskip}
\hline
\end{tabular}
\smallskip\\
Notes:
\smallskip\\
'-99.9' indicates that the object is not detected in the corresponding band or that the magnitude could not be measured.\\
'-88.8' indicates that the object is out of the boundaries of the image obtained in the corresponding band.\\
\end{table}
\end{landscape}

\begin{landscape}
\begin{table}
\caption{\label{YSOs_ECX6-21}Young stellar objects in ECX 6-21}
\begin{tabular}{lccccccccccccccccc}
\hline\hline
\noalign{\smallskip}
Object No. & $\alpha (2000)$ & $\delta(2000)$ & Class & $g'$ & $r'$ & $i'$ & $J$ & $H$ & $K_S$ & $I1$ & $I2$ & $I3$ & $I4$ & $M ({\rm M}_\odot)$ & $L$ (L$_\odot$) & $A_K$ & $n$ \\
\noalign{\smallskip}\hline\noalign{\smallskip}
\noalign{\medskip \centerline{ECX~6-21} \medskip}
ECX 6-21 01 & 20:30:18.13 & 41:16:00.5 & II    & -99.9 & -99.9 & -99.9 & 19.37 & 16.65 & 15.30 & 13.29 & 12.79 & -99.9 & -99.9 &  0.4 & -0.05 & 2.26 & -1.3 \\  
ECX 6-21 02 & 20:30:25.24 & 41:16:00.1 & II    & -99.9 & -99.9 & -99.9 & -99.9 & -99.9 & 15.54 & 13.61 & 13.14 & 12.69 & -99.9 &  0.3 & -0.4 & 2.00 & -1.3 \\  
ECX 6-21 03 & 20:30:27.29 & 41:15:27.7 & Trans & 21.20 & 17.88 & 15.87 & 11.69 & 10.27 &  9.62 &  9.62 &  9.13 &  8.20 & -99.9 &  3.7 &  1.9 & 0.88 & -2.6 \\  
ECX 6-21 04 & 20:30:27.34 & 41:16:06.3 & I     & -99.9 & -99.9 & -99.9 & -99.9 & -99.9 & 14.74 & 10.89 &  8.99 & -99.9 & -99.9 &  0.5 &  0.1 & 2.00 & 2.8  \\  
ECX 6-21 05 & 20:30:27.37 & 41:16:00.1 & FS    & -99.9 & -99.9 & -99.9 & -99.9 & 14.85 & 12.84 &  9.95 &  9.20 & -99.9 & -99.9 &  3.7 &  1.9 & 3.22 & -0.5 \\  
ECX 6-21 06 & 20:30:27.64 & 41:15:33.2 & Trans & -99.9 & 21.07 & 18.46 & 13.41 & 11.44 & 10.36 &  9.44 &  8.73 &  7.96 & -99.9 &  3.7 &  1.9 & 1.33 & -1.7 \\  
ECX 6-21 07 & 20:30:28.12 & 41:15:05.2 & II    & -99.9 & -99.9 & -99.9 & 17.03 & 15.52 & 14.81 & 13.90 & 13.18 & -99.9 & -99.9 &  0.2 & -0.6 & 0.69 & -1.5 \\  
ECX 6-21 08 & 20:30:28.75 & 41:15:18.0 & II    & -99.9 & -99.9 & -99.9 & 18.57 & 16.74 & 15.60 & 13.38 & 13.06 & -99.9 & -99.9 &  0.2 & -0.6 & 1.47 & -0.6 \\  
ECX 6-21 09 & 20:30:28.75 & 41:15:58.8 & I     & -99.9 & -99.9 & -99.9 & -99.9 & -99.9 & -99.9 & 10.46 &  8.57 &  6.99 & -99.9 &      &      &      &      \\
ECX 6-21 10 & 20:30:29.16 & 41:15:28.9 & II    & -99.9 & -99.9 & -99.9 & 18.40 & 14.93 & 12.94 & 10.48 &  9.89 &  9.10 &  8.52 &  3.7 &  1.8 & 3.40 & -1.2 \\  
ECX 6-21 11 & 20:30:29.38 & 41:15:57.7 & I     & -99.9 & -99.9 & -99.9 & -99.9 & -99.9 & 13.14 &  9.49 &  8.44 &  7.19 & -99.9 &  2.8 &  1.1 & 2.00 &  1.5 \\ 
ECX 6-21 12 & 20:30:32.31 & 41:16:19.2 & Trans & -99.9 & 20.68 & 18.56 & 14.68 & 13.29 & 12.74 & 12.27 & 12.25 & 11.79 & 10.64 &  1.7 &  0.7 & 0.93 & -2.2 \\ 
ECX 6-21 13 & 20:30:34.16 & 41:16:28.7 & Trans & -99.9 & -99.9 & -99.9 & 14.66 & 12.88 & 11.89 & 10.59 &  9.97 &  9.82 &  9.09 &  4.0 &  2.2 & 2.04 & -1.9 \\ 
ECX 6-21 14 & 20:30:34.80 & 41:15:33.2 & Trans & 17.34 & 14.80 & 13.42 & 11.01 & 10.33 & 10.03 & 10.10 &  9.85 &  9.85 &  8.93 &  3.6 &  1.8 & 0.53 & -2.7 \\
\noalign{\medskip \centerline{ECX 6-21-W} \medskip}
ECX 6-21-W 01 & 20:29:49.56 & 41:16:30.8 & FS    & -88.8 & -88.8 & -88.8 & 18.96 & 17.77 & 17.18 & 16.41 & 16.43 & 14.26 & 12.84 & 0.07 & -1.4 & 0.78 & -0.3 \\  
ECX 6-21-W 02 & 20:29:55.66 & 41:16:29.1 & Trans & -88.8 & -88.8 & -88.8 & 15.40 & 13.44 & 12.26 & 10.67 & 10.35 &  9.83 &  9.06 & 4.2 &  2.5 &  2.33 & -1.8 \\  
ECX 6-21-W 03 & 20:29:58.53 & 41:16:41.0 & II    & -88.8 & -88.8 & -88.8 & -99.9 & 18.22 & 15.61 & 12.55 & 11.28 & 10.45 & -99.9 & 3.3 &  1.4 &  5.10 & -0.9 \\  
ECX 6-21-W 04 & 20:29:59.12 & 41:16:47.2 & Trans & -88.8 & -88.8 & -88.8 & -99.9 & 14.32 & 12.10 & 10.23 & 10.09 &  9.58 & -99.9 & 3.7 &  1.9 &  2.75 & -2.1 \\   
ECX 6-21-W 05 & 20:30:00.40 & 41:16:10.8 & I     & -88.8 & -88.8 & -88.8 & 18.26 & 16.87 & 16.14 & 14.60 & 14.92 & -99.9 & 10.53 & 0.1 & -1.2 &  0.66 &  0.8 \\  
\noalign{\medskip \centerline{ECX 6-21-S} \medskip}
ECX 6-21-S 01 & 20:30:32.92 & 41:11:06.0 & I 	 & -99.9 & -99.9 & -99.9 & -99.9 & -99.9 & 17.08 & 14.13 & 12.97 & 11.84 & 10.89  & 0.1 & -1.2 &  2.00 &  0.6 \\  
ECX 6-21-S 02 & 20:30:33.52 & 41:11:05.5 & II	 & -99.9 & -99.9 & -99.9 & -99.9 & -99.9 & 16.20 & 14.66 & 14.36 & 13.89 & -99.9  & 0.2 & -0.6 &  2.00 & -1.9 \\  
ECX 6-21-S 03 & 20:30:34.33 & 41:11:01.4 & II	 & 22.00 & 20.64 & 18.87 & 16.79 & 16.18 & 15.90 & 15.26 & 14.34 & -99.9 & -99.9  & 0.1 & -1.2 &  0.03 & -1.2 \\  
ECX 6-21-S 04 & 20:30:34.39 & 41:10:56.2 & II	 & -99.9 & -99.9 & -99.9 & -99.9 & 17.28 & 15.90 & 14.52 & 14.00 & -99.9 & 10.47  & 0.2 & -0.5 &  2.04 & -0.2 \\  
ECX 6-21-S 05 & 20:30:36.09 & 41:11:15.0 & I 	 & -99.9 & -99.9 & -99.9 & -99.9 & -99.9 & 16.78 & 14.14 & 13.28 & 12.63 & -99.9  & 0.2 & -0.6 &  2.00 & -0.1 \\
\noalign{\medskip}
\hline
\end{tabular}
\smallskip\\
Notes:
\smallskip\\
'-99.9' indicates that the object is not detected in the corresponding band or that the magnitude could not be measured.\\
'-88.8' indicates that the object is out of the boundaries of the image obtained in the corresponding band.\\
\end{table}
\end{landscape}

\end{appendix}
\end{document}